# Computer modeling of natural silicate melts: what can we learn from *ab initio* simulations.

Rodolphe Vuilleumier[1], Nicolas Sator and Bertrand Guillot

*Laboratoire de Physique Théorique de la Matière Condensée,*

*Université Pierre et Marie Curie (Paris 6), UMR CNRS 7600, case courrier 121, 4 place jussieu,*

*75252 Paris cedex 05, France.*

The structural and dynamical properties of four silicate liquids (silica, rhyolite, a model basalt and enstatite) are evaluated by *ab initio* molecular dynamics simulation using the density functional theory and are compared with classical simulations using a simple empirical force field. For a given composition, the structural parameters of the simulated melt vary little between the two calculations (*ab initio* versus empirical) and are in satisfactory agreement with structure data available in the literature. In contrast, ionic diffusivities and atomic vibration motions are found to be more sensitive to the details of the interactions. Furthermore, it is pointed out that the electronic polarization, as evaluated by the ab initio calculation, contributes significantly to the intensity of the infrared absorption spectra of molten silicates, a spectral feature which cannot be reproduced using nonpolarizable force field. However the vibration modes of $TO_4$ species and some structural details are not accurately reproduced by our ab initio calculation, shortcomings which need to be improved in the future.

---

[1] Present address: Laboratoire PASTEUR, UMR8640, Département de Chimie, Ecole Normale Supérieure, 75005, Paris, France

# 1. Introduction

Although today the volume of melt involved in the production of oceanic crust at mid-ocean ridges (~21 km$^3$/y, see Perfit and Davidson, 2000), in oceanic island magmatism (~4km$^3$/y) and in volcanism above the subduction zones (~9km$^3$/y) is small with respect to the amount of minerals present in the mantle, the generation and migration of melts are at the origin of the chemical differentiation of our planet. Thus, the partitioning of elements between liquids and minerals, the migration of melts through the crystalline matrix or the mobilization of volatiles from the source regions (Dasgupta and Hirschmann, 2006; Asimow and Langmuir, 2003), are some of the key mechanisms at work in the Earth's mantle requiring some theoretical guidance to be better understood. In this context, the modeling of silicate liquids by molecular simulation is a very useful tool to describe many details of the above mechanisms at the microscopic (atomic) level.

Classical molecular dynamics simulations based on empirical interatomic potentials (EPMD) are now widely used to simulate molten silicates of increasing complexity (e.g. Guillot and Sator, 2007a, and references therein). The advantage of using empirical potentials is that they are easy to implement in simulation codes and are rather inexpensive in computer time with the result that a rather large number of atoms (N~$10^3$-$10^4$) can be simulated over long period of time (t~10 ns). The disadvantage is that the quality of the results crucially depends on the accuracy of the force field used as input. The advent of the density functional theory (DFT) in quantum chemical calculations (Hohenberg and Kohn, 1964; Kohn and Sham, 1965) and its implementation in a molecular dynamics scheme by Car and Parrinello (1985) has been a breakthrough for condensed matter physics. Newton's equations of classical mechanics are solved iteratively to describe the motions of individual atoms in the sample and forces acting on those at each time-step are obtained by an electronic structure calculation at the density functional theory level. This is in contrast to classical MD simulations, where forces are given by an empirical interaction potential usually fitted to reproduce some experimental data. The challenge with a magma composed of nine major oxides is that the ionicity-covalency ambivalence of the atomic interactions may change not only with the composition of the melt but also transiently at the mercy of the chemical disorder induced by the diffusion of the atoms into the melt. The strength of the first-principle approach (AIMD) resides in the explicit treatment of the electrons that allows for the description of matter in many different conditions of temperature, pressure and composition, without any fitted parameter. However, the very high computational cost (3 to 4 orders of magnitude more expensive in computer time than classical simulations) restricts its use to small systems (N~100 atoms) evolving over short period of time (t~10 ps). Thus, some studies are devoted to the development of more accurate interatomic potentials for silicate materials by using AIMD calculations as benchmarks. For instance, improved interatomic potentials based upon ab initio calculations have been proposed for $SiO_2$ (Tangney and Scandolo, 2002), MgO (Aguado et al., 2002; Tangney and Scandolo, 2003; Aguado and Madden, 2004), $Al_2O_3$ (Jahn et al., 2006), the alkaline earth oxides CaO, SrO and BaO ( Aguado et al., 2003), and the CMAS system (Jahn and Madden, 2007; Adjaoud et al., 2008). However, the price to pay for these improvements is a greater analytical complexity of the potential due to its nonpairwise additivity (accounting for polarization effects) which increases drastically the computer time (generally by more than a factor of 10). Moreover, if the AIMD calculations show some weaknesses in the description of the investigated system or if the analytical form of the potential does not fit well with the quantum force field, these flaws will be incorporated into the fitting procedure. Hence, when investigating a given material by AIMD calculation, it is important to check the level of accuracy by comparing the results with the experimental data available for this substance but also to compare the results with those obtained from classical simulations performed with empirical potentials used in the literature.

Here we have investigated by AIMD simulation four silicate liquids, representative of the composition range encountered with natural magmas namely, pure silica, an iron-free rhyolite, a model basalt (the anorthite-diopside eutectic: 36mol%-64mol%), and an ultrabasic melt of enstatitic composition ($MgSiO_3$). The structure, ionic diffusivities and infrared absorption spectra of the simulated melts are compared with their counterparts evaluated by classical MD simulations using an empirical potential developed recently for the $K_2O$-$Na_2O$-$CaO$-$MgO$-$FeO$-$Al_2O_3$-$TiO_2$-$SiO_2$ system (Guillot and Sator, 2007a,b). A detailed comparison with experimental data in the literature for melts of similar composition is given.

## 2. Computational details

Ab initio molecular dynamics simulations were performed with the CPMD code (Hutter et al., 1995) in the microcanonical ensemble using the Kohn-Sham formulation of the density functional theory in the generalized gradient approximation BLYP (Becke, 1988; Lee et al., 1988). Core electrons were not treated explicitly and norm conserving plane wave pseudopotentials of the Trouiller-Martins type were used for all atoms (Trouiller and Martins, 1991) except for aluminum for which a similar Bachelet-Hamann-Schlüter type pseudopotential was employed (Bachelet et al. 1982) in agreement with earlier calculations on alumina (Holm et al., 1999). For the cations, Na and K were described by semicore pseudopotentials, while to minimize the computational cost only two valence electrons were included for Ca and Mg. For Mg it was essential to include nonlinear core corrections to reproduce the Mg-O distance (Louie et al. 1982). All pseudopotentials were transformed to non-local form following the method of Kleinman and Bylander (Kleinman and Bylander 1982), up to $d$ angular momentum for Si, Al and Ca, and $p$ angular momentum for the remaining species. Kohn-Sham orbitals were expanded in plane waves with a kinetic energy cut off of 70 Ry. The Newton equations were solved with a fictitious mass $\mu = 300$ a.u. for the electrons and a time step of 0.048 fs. The simulation cell of constant volume (microcanonical ensemble) was periodically replicated and contained about 96~100 atoms according to the composition of the silicate melt under investigation (see Table 1). The length of the simulation runs were typically 6~7ps the first 1~2ps of which were discarded.

The initial configurations for AIMD simulations were prepared from fully equilibrated runs generated by classical MD simulations based upon the empirical potential (EPMD) developed for the nine component system $K_2O$-$Na_2O$-$CaO$-$MgO$-$FeO$-$Fe_2O_3$-$Al_2O_3$-$TiO_2$-$SiO_2$ (for the potential parameters see Table 2 in Guillot and Sator, 2007a). These EPMD simulations were performed with the DL-POLY code (Smith and Forrester, 1996), for system sizes identical to those used in AIMD simulations (i.e. N~100 atoms) for a proper comparison and some runs were also performed with a larger system (N~1000) to evaluate finite size effects.

The equations of motion for the ions were solved with a time step of 1fs. The long range coulombic interactions (periodic boundary conditions) were accounted for by a Ewald sum with $\alpha L$~6, where $\alpha$ is an optimized value of the width of the charge distribution on each ion and L the size of the unit cell. As usual with periodic boundary conditions, long range correction terms were added to the potential energy and the pressure to account for the dispersion forces between atoms for distances beyond the minimum image convention (i.e. for d>L/2, where L is the length of the simulated cell). Long simulation runs (1~100ns or $10^6$~$10^8$ time steps) were performed in the microcanonical ensemble (NVE) and in the isobaric-isothermal ensemble (NPT) to cross check the statistical uncertainties of the evaluated properties. The simulations were performed near zero pressure and at a high enough temperature (T~2200-2500K) to be sure that we were exploring the liquid phase for all composition. This was verified by analyzing the evolution of the mean square displacements of the atoms with the simulation time. However it is worth noting that the statistical fluctuations on pressure (NVE) or density (NPT) are rather large for a system size of ~100 atoms. When imposing a zero pressure (NPT), the density of the melt fluctuates by ±4%, although its running average is much better estimated (~±1%). In the present case at 2273 K and P~0, the density of the simulated melt is 2.25±0.09 $g/cm^3$ for $SiO_2$, 2.23±0.10 $g/cm^3$ for rhyolite, 2.54±0.10 $g/cm^3$ for $An_{36}Di_{64}$ and 2.56±0.09 $g/cm^3$ for enstatite. These values compare very well with data in the literature based upon direct measurements or obtained from multicomponent linear least squares regressions of molar volume. However, in the experimental literature, the density of silicate liquids is measured by the Archimedean method for a temperature generally not exceeding ~2000 K and hence an extrapolation has to be done to reach the temperature range (2200-2500 K) that we investigate by simulation. Thus for enstatite at 2273K the data of Courtial and Dingwell (1999) lead to a density about 2.49 g/cm$^3$ whereas for An$_{36}$Di$_{64}$ the data of Taniguchi (1989) lead to 2.55 g/cm$^3$ after a linear extrapolation up to 2273 K. For rhyolite, the multicomponent regression formula of Lange and Carmichaël (1987) leads to a density of 2.23 g/cm$^3$, the recent one of Ghiorso and Kress (2004) gives a value close to 2.20 g/cm$^3$ when the density measurements of Knoche et al. (1995) lead to 2.26 g/cm$^3$ for haplogranitic compositions but after a large extrapolation between 1233K and 2273 K. based on the thermal expansivity. For silica, the situation is more controversial as emphasized in a review by Hudon et al. (2002) where a value near 2.20 g/cm$^3$ at 2273 K is recommended.

Because the AIMD calculations are expensive in computer time, it is convenient to optimize the process of equilibration by starting from atomic configurations which are representative of the silicate melts under investigation. In starting from a well equilibrated configuration generated by EPMD simulation, we have checked that the AIMD calculation tends to equilibrate rapidly toward an energetic basin very close to the one sampled by the EPMD calculation. This is illustrated in Fig.1 for the anorthite-diopside melt where the running average of the potential energy and the temperature are shown as function of time: the equilibrium is reached rapidly, after about 1.5 ps (for more information see the legend of the figure). But, with production runs of only 5ps for AIMD simulations, it is not possible to check with accuracy if the density chosen for the melt corresponds to a near zero pressure in the simulation. Moreover, it is known that the use of the generalized gradient approximation (BLYP functional) for the exchange correlation energy tends to underestimate by ~2-3% the density of silicate minerals when the use of the local density approximation (LDA), although less rigorous, leads to a slightly better estimation (Oganov et al., 2001, 2004; Karki et al., 2006; Wan et al., 2007). But GGA BLYP remedies some of the weaknesses of LDA, and especially the tendency exhibited by the latter one to badly describe the evolution of the Si coordination number with pressure (Hamman, 1996). In fact the slight underestimation of the density when using GGA BLYP originates from the dispersion energy between atoms which is not correctly evaluated (Wu et al., 2001; Wu and Yang, 2002).

In practice, the production runs (AIMD and EPMD) were performed in the microcanonical ensemble with the experimental values of the density at 2273K as discussed above, i.e. 2.20 g/cm$^3$ for $SiO_2$ (Hudon et al., 2002), 2.26 g/cm$^3$ for rhyolite (Knoche et al., 1995), 2.55g/cm$^3$ for $An_{36}Di_{64}$ (Taniguchi, 1989) and 2.49 g/cm$^3$ for enstatite (Courtial and Dingwell, 1999), knowing that in EPMD simulations, the pressure of the four silicate melts fluctuates around zero pressure for these densities.

# 3. Properties of the simulated silicate melts: ab initio versus empirical force field.

## 3.1 Comparison with X-ray diffraction data.

From the atomic configurations generated by AIMD and EPMD simulations we have evaluated the atom-atom pair distribution functions (PDF) in the four silicate liquids (see Figs.2 and 3). These PDFs can be combined with each other to evaluate for instance the radial distribution function (RDF) observed in X-ray diffraction studies. Lately much interest has been devoted to fast X-ray scattering measurements on high temperature liquids in using aerodynamic levitation and laser heating (Krishnan and Price, 2000). The use of high energy X-ray sources gives access to the local structure in the liquid with an unprecedented accuracy. Thus, liquid silica (Mei et al., 2007), liquid aluminates (Hennet et al., 2007; Mei et al., 2008) and magnesium silicate liquids (Wilding et al., 2008) have been investigated by this technique.

In an X-ray diffraction experiment, the RDF is calculated from the k-dependent structure factor $S(k)$ by the equation,

$$G_X(R) = \frac{1}{2\pi^2 R \rho_0} \int_0^\infty k S(k) \sin(kR)\, dk \qquad (1)$$

where $\rho_0 = N/V$ is the atomic density of the melt and $S(k)$ the structure factor defined by,

$$S(k) = \rho_0 \sum_{i,j}^N C_i\, C_j\, f_{ij}\,(k) \int_0^\infty R(g_{ij}(R) - 1) \frac{\sin(kR)}{k}\, dR. \qquad (2)$$

In the above equation, $C_i$ is the mole fraction of species $i$, $g_{ij}(R)$ is the pair distribution function between atoms $i$ and $j$, and $f_{ij}(k)$ is defined from the tabulated (Waasmaier and Kirfel, 1995) k-dependent atomic scattering factors $f_i(k)$ by the relation,

$$f_{ij}(k) = f_i(k)\, f_j(k)\, / \left[\sum_{i=1}^N C_i\ f_i(k)\right]^2 \qquad (3)$$

In Fig.4 is compared the X-ray RDF for liquid $SiO_2$ at 2373K (Mei et al., 2007) with those obtained by AIMD and EPMD calculations. A good agreement is found between experimental data and computer generated $G_X(R)$, although the EPMD calculation seems to be a little bit superior to the AIMD calculation in reproducing more accurately the detail of the experimental curve in the 2~3.5 A range. However, it is noteworthy that the AIMD simulation run is too short (5ps) to sample a fully relaxed silica melt because of the very slow diffusivity of Si and O atoms at the investigated temperature (see section 3.3). To quantify the evolution of $G_X(R)$ with the length of the simulation run, we have performed by EPMD several simulation runs at 5 ps, 10 ns ( 1 ns = 1,000 ps) and 100 ns, respectively, knowing that a full diffusion of the atoms is observed in the simulated melt at 2273K only after a few nanoseconds. The results indicate that between 5 ps and 10 ns, $G_X(R)$ becomes slightly more structured (only the run at 10 ns is shown in Fig.2) and there is no more evolution between 10 and 100 ns. Hence, if we were able to perform a much longer simulation run by AIMD (10 ns instead of 5 ps, a simulation time unreachable with our present computational resources), the resulting $G_X(R)$ would be slightly more structured than the one presented in Fig.2 and likely very close to the curve obtained by EPMD. To be complete, notice also that there is no detectable difference on $G_x(R)$ in increasing the number of simulated atoms from ~100 to ~1,000 in EPMD calculations.

The decomposition of the RDF in Si-O, Si-Si and O-O components shows that the Si-O distance is equal to 1.63 A by EPMD calculation and 1.65 A by AIMD when Mei et al. (2007) evaluate this distance at 1.626±0.005 A in the liquid at 2373K. The latter value is in agreement with an earlier estimation (~1.63 A) based upon the thermal expansion of the Si-O bond measured by neutron scattering experiment (Tucker et al., 2000). The second and the third peak of the RDF are clearly visible and are dominated by the first neighbor peaks associated with $g_{OO}(R)$ and $g_{SiSi}(R)$, respectively, whereas the fourth peak around 4.3 A is the combination of the Si-O, Si-Si and O-O second neighbors (see also Fig.2). On the contrary to the first three peaks which fit well the experimental data, the fourth peak is shifted in the calculations by +0.15 A with respect to the experimental one ($R_{max}$~4.15 A). In the following we will discuss the consequence of this deviation on the inter-tetrahedral Si-O-Si bond angle distribution.

In the case of rhyolite, at our knowledge, there is no X-ray diffraction data available in the high temperature liquid. Consequently, we have compared in Fig.5 our calculated RDFs with the X-ray data of liquid albite ($NaAlSi_3O_8$) measured by Sugiyama et al. (1996). Indeed, the effect of the difference in composition between rhyolite and albite (Na/Si = Al/Si = 0.33 in albite and (Na+K)/Si = Al/Si = 0.19 in rhyolite) is assumed to be small on $G_x(R)$ in considering the close similarity between X-ray data of rhyolitic glasses (Zotov et al., 1989; Okuno et al., 1996) and those of the albite glass (Hochella and Brown, 1984). Moreover it has been shown that for silicic glasses and melts, the difference in temperature between the glass and the liquid has only a weak influence on the X-ray RDF (Taylor et al., 1980; Okuno and Marumo, 1982; Mei et al., 2007), except for the height of the Si-O contribution which decreases when the temperature increases. Coming back to Fig.5, the overall agreement between experiment and computer simulations is very good, and especially with the EPMD calculation which reproduces accurately the shape of the second and the third peak of the RDF for liquid albite. The decomposition of the RDF in atom-atom contributions shows that the first peak is dominated by the Si-O first neighbor distance (1.64 A with EPMD, 1.65 A with AIMD) with a weak contribution of Al-O distance (1.73 A with EPMD, 1.77A with AIMD). The second, third and fourth peaks are governed by the O-O and Si-Si first neighbor distances and the Si-O second neighbor distance, respectively. Notice that Na-O and K-O distances (around 2.40-2.46 A and 2.80-3.20 A, see Table 2) contribute very little to the RDF, less than 1% of the Si-O contribution, and are virtually undetectable on the RDF. Furthermore, only Si-Al distances give rise to a small but non negligible contribution around 3.2 A, whereas above 2.5 A the contributions of Si-Na, Si-K and Al-Al distances are vanishingly small. As in the case of liquid silica, we observe a slight discrepancy in the position of the fourth peak which is shifted at a greater distance in our simulations (4.3 A instead of 4.15 A in liquid albite). Finally, it is intriguing that the small peak around 2.1 A exhibited by the experimental RDF of liquid albite does not correspond to a remarkable distance in the simulated melt and is likely an artifact coming from the Fourier transform of S(k) (a similar feature is also found in the experimental RDF of liquid silica).

The X-ray RDF for the $An_{36}Di_{64}$ eutectic liquid simulated by AIMD and EPMD are presented in Fig.6. The two calculated functions are very similar to each other and present essentially two main peaks at 1.65 A and 3.1 A, respectively. The first peak is asymmetric and is produced by the combination of Si-O and Al-O first neighbor distances (see Table 2 for Si-O and Al-O values), whereas the second peak is broad and structured and is generated by the overlap of Ca-O, O-O, Si-Si, Si-Al and Mg-Mg first neighbor contributions. Thus the Ca-O distance (~2.4-2.5 A, see Table 2) gives a weak shoulder on the low R-flank of the second peak, the O-O contribution makes a hump near 2.65 A while the Si-Si and Si-Al inter-tetrahedral distances enhance the intensity of the peak near 3.1 A (the other cation-cation correlations, although effective in this R range, contribute very little to the RDF). The Mg-O contribution ($R_{max}$~2.0 A) is hardly detectable since its main effect is to fill in partially the dip located between the two main peaks. At larger distances, a bump is visible on the RDF near 4.2 A, a feature coming from the overlap of the Si-O second neighbor contribution with O-O and cation-cation correlations.

Considering that X-ray diffraction data for the $An_{36}Di_{64}$ eutectic liquid or for a basic melt of similar composition are lacking yet in the experimental literature, we have compared our results with X-ray diffraction data for pyrope ($Mg_3Al_2Si_3O_{12}$) and grossular ($Ca_3Al_2Si_3O_{12}$) glasses (Okuno and Marumo, 1993). Thus the peaks observed at 1.68, 2.1, 2.6, 3.2 and 4.2 A in pyrope glass and at 1.7, 2.4, 3.1, 3.6 and 4.2 A in grossular glass are in excellent agreement with those contributing to the RDF of our simulated anorthite-diopside melt (with the following assignments: $R_{Si-O}$ + $R_{Al-O}$~1.7 A, $R_{Mg-O}$~2.1 A, $R_{Ca-O}$~2.4 A, $R_{O-O}$~2.6 A, $R_{Si-Si}$~3.1 A, $R_{Si-Ca}$~3.6 A, and $R_{(Si-O)_2}$~4.2 A).

Recently, accurate X-ray diffraction data on magnesium silicate liquids at high temperature (~2273K) have been obtained by combining high energy X-rays, aerodynamic levitation and laser heating (Wilding et al., 2008). In Fig.7 are compared the X-ray RDF for liquid enstatite with our results by AIMD and EPMD simulations. Notice that the small variations observed between the pair distribution functions $g_{ij}(R)$ generated by AIMD and EPMD (see Figs.2 and 3) are slightly amplified in the final RDFs (see Fig.7). Although the RDF generated by EPMD is slightly more structured than the one generated by AIMD, it is difficult to settle which one is the most accurate to reproduce the experimental function for liquid enstatite. With regard to the assignment of peaks, our results confirm the conclusions reached by the experimental study of Wilding et al. (2008). Thus the first peak is centered on the Si-O distance at about 1.63-1.65 A and is characterized by a shoulder near 2.1 A which is the signature of the Mg-O distance at 2.0 A. A hump is visible near 2.6-2.7 A which corresponds to the O-O first neighbor distance, whereas the peak at 3.1 A is produced by the overlap of Si-Si, Si-Mg and Mg-Mg first neighbor distances. At larger distance a third peak is visible on the experimental

RDF near 4.05 A (~4.3 A in the simulations), a feature whose the main origin is the Si-O second neighbor distance (in addition to other atom-atom contributions, see Fig.7).

## 3.2. Structural parameters in silicate melts.

In order to be more quantitative about the distribution of oxygens around the cations we have evaluated the abundance of n-coordinate species and their corresponding average cation-oxygen distance. In the liquid phase, the interatomic distance corresponding to the maximum of the first peak in the cation-oxygen PDF does not coincide with the cation-oxygen mean distance of the most abundant n-coordinated species. As shown in Table 3, [4]Si is by far the most abundant n-coordinated Si species in the four melts investigated here, with a mean [4]Si-O distance of 1.69-1.70A in AIMD calculations and about 1.67-1.68A in EPMD (see Fig.8) when the nearest neighbor distance given by the first maximum of the corresponding PDF is 1.65A and 1.63A respectively. Nevertheless, these Si-O distances vary little with melt composition (at most 0.01A), a finding in agreement with neutron diffraction data in sodium silicate glasses (Wright et al., 1991) but at variance with structure data in jadeite and nepheline glasses where the Si-O first neighbor distance increases more significantly with the ratio Al/(Al+Si) (Taylor and Brown, 1979a). As for [5]Si species, it is found in a very low abundance in silica and rhyolite (1% or less, see Table 3) but becomes appreciable in $An_{36}Di_{64}$ and enstatite (3.4-4.7%) with a mean [5]Si-O distance about 1.81-1.84 A (see Fig.8). From the experimental viewpoint, 5-coordinated Si species has been detected in alkali silicate glasses at ambient pressure but in a much lower proportion (≤0.05% in Stebbins, 1991; Xue et al., 1991). The fact that we find [5]Si in a much higher abundance in our simulations (both in AIMD and EPMD) could suggest that the high temperature liquid promotes the occurrence of 5- coordinated species.

A similar situation occurs with [5]Al which is found in a significant amount both in the rhyolitic melt (20.0% by AIMD and 9.9% by EPMD) and in the anorthite-diopside eutectic liquid (35.8% by AIMD and 22.8% by EPMD), the more depolymerized the melt the higher the abundance of [5]Al. Notice that [6]Al is found in a much smaller proportion in both melts (~0.4-2%). Evidence of [5]Al (and [6]Al as well, but in a much smaller proportion $\leq 1\%$) is well documented in aluminosilicate glasses by $^{27}$Al NMR (Stebbins et al., 2000; Toplis et al., 2000; Neuville et al., 2004 ,2006, 2007, 2008a) but its abundance is lower than the one found in our simulations at liquid temperature; e.g. about 11% in $Mg_{1.5}Ca_{1.5}Al_2Si_6O_{18}$ (Kelsey et al., 2008) and about 10% in Mg/Ca aluminosilicate glasses of composition close to $An_{36}Di_{64}$ (see Fig.6 in Neuville et al., 2008a). Although a clear view of the influence of the temperature effects on the coordination of Al in aluminosilicate glasses to bridge the gap with the liquid phase is still lacking, some recent advances have been made by investigating glass samples by high-resolution NMR spectroscopy at various quench rates (Kiczenski et al., 2005; Stebbins et al., 2008). Thus Stebbins et al (2008) predicts a [5]Al/[4]Al ratio about 30% in CAS liquids at 2000 °C, a value quite compatible with our evaluations for $An_{36}Di_{64}$. More generally, the significant abundance of [5]Al in silicate liquids predicted many years ago by classical MD simulations (e.g. Scamehorn and Angell, 1991; Poe et al., 1992a; Stein and Spera, 1995; Nevins and Spera, 1998; Morgan and Spera, 2001; Winkler et al., 2004; Guillot and Sator, 2007a) now appears to be credible at the light of the aforementioned experimental investigations. The fact that our ab initio calculations also predict the presence of a significant concentration of [5]Al in felsic and basic melts merits to be emphasized in this context.

With regard to the distribution of oxygen around the $Mg^{2+}$ cation, this is still a controversial issue. In some studies on diopside and enstatite glasses it was concluded that Mg was fourfold coordinated (Tabira, 1996; Wilding et al., 2004a,b), fivefold coordinated in others (Ildefonse et al., 1995; Li et al., 1999) or even six fold (Kroecker and Stebbins, 2000). A recent study of enstatite and calcomagnesian silicate glasses by ultra-high field $^{25}$Mg NMR spectroscopy (Shimoda et al., 2007) points out the occurrence of multiple Mg sites with several degrees of distortion, the octahedral sites being favored. In our simulations (Table 3),

Mg is distributed mainly through fourfold, fivefold and six fold coordination with a mean coordination number around 5.2. These results agree with a study of MAS glasses using X-ray and neutron diffraction data combined with reverse Monte Carlo analysis (Guignard and Cormier, 2008). Moreover, it is worthwhile to emphasize that [4,5,6]Mg-O distances evaluated by simulation in $An_{36}Di_{64}$ and enstatite ($R_{[4,5,6]}$~2.12-2.34 A, see Fig.9) are significantly larger than the nearest neighbor distance (~2.0-2.1 A) characterizing the first maximum of the Mg-O PDF as observed in diffraction studies (see Okuno and Marumo, 1993; Taniguchi et al., 1995, 1997; Kohara et al., 2004; Guignard and Cormier, 2008; Wilding et al., 2008).

The environment around calcium atoms in the basic liquid ($An_{36}Di_{64}$) is characterized by a broad distribution of n-coordinated species with n=6-10 (average coordination number ~7.9-8.6, see Tables 2 and 3) and [n]Ca-O distances in the range 2.65-2.85A (Fig.9). This is in agreement with the broad distribution of Ca-O distances found in liquid anorthite by X-ray absorption spectroscopy (Neuville et al., 2008b). Moreover, a study of a CMAS amorphous slag by NMR spectroscopy on $^{43}Ca$ (Shimoda et al., 2006) has pointed out 5 different sites which are assigned to 6-, 7-, and 8- coordinated species by analogy with Ca environments found in Ca-bearing solid phases (Dupree et al., 1997). Thus as long as the liquid phase is concerned, the present simulations suggest that the range of Ca-O distances covered by the first coordination shell is greater than expected from a simple observation of diffraction data for which the Ca-O nearest neighbor distance in Ca- bearing silicate glasses evolves in a limited range (~2.40-2.45 A). As a matter of fact, a rapid survey of the X-ray diffraction data of the literature shows that the Ca-O first neighbour distance lies in the range 2.35-2.42 A in wollastonite glass and melt (Eckersley et al., 1988; Waseda and Toguri, 1977, 1990; Taniguchi et al., 1997), 2.32-2.40 A in calcium aluminosilicate glasses (Okuno and Marumo, 1993; Cormier et al., 2000; Petkov et al; 2000), 2.45 A in diopside glass (Taniguchi et al., 1995) and 2.40 A in liquid aluminate (Hennet at al., 2008). By comparison we obtain 2.38 A (by EPMD) and 2.50 A (by AIMD) in the anorthite-diopside eutectic.

In the case of the alkali cations in our simulated rhyolitic liquid, the sodium-oxygen and potassium-oxygen PDFs are broad and asymmetric (see Fig.3). These features are the structural counterpart of the large difference in diffusivity between alkali ions and network former ions (see the next section, and for a discussion about the relationship between atomic structure and ionic transport in oxide glasses, see Greaves and Ngai, 1995). The Na-O nearest neighbor distance is equal to 2.40 A in AIMD and 2.46 A in EPMD calculations, as compared with 2.40-2.45 A in sodium silicate glasses (Zotov and Keppler, 1998). Furthermore, the distribution of oxygens around sodium shows a bell shaped curve (Table 3) centered about 7~8-coordinated species corresponding to Na-O distances about 2.7-3.0 A in AIMD and 2.8-3.0 A in EPMD calculations. By comparison, a high field $^{23}$Na NMR study (Lee and Stebbins, 2003) of sodium aluminosilicate glasses leads to a distribution of Na-O distances in Si-rich glasses quite comparable (~2.6-3.0 A) to what we obtained. As for the K-O nearest neighbor distance, it is equal to 2.8 A with AIMD and 3.2 A with EPMD, whereas the first coordination shell extends to 4.0-4.2 A. These results and especially those obtained by AIMD are in agreement with a neutron diffraction study of a potassium disilicate melt at 1350K (Majérus et al., 2004) where the potassium-oxygen PDF shows a broad maximum around 2.7-3.2A followed by a second peak at 5.A, the first shell around the potassium containing about 9-10 oxygen atoms (~9 in our calculations, see Table 2).

The oxygen-oxygen mean distance is found to be practically invariant in the range of composition investigated here (2.65-2.66A by EPMD and 2.70-2.72A by AIMD, see Fig.2 and Table 2). Neutron and X-ray diffraction data show that, indeed, this distance lies in the range 2.65-2.69A for a large variety of silicate glasses and melts of low Al contents (Waseda and Suito, 1977; Waseda and Toguri, 1977, 1990; Zotov et al., 1989; Petkov et al., 2000; Mei et al., 2007; Wilding et al., 2008; Guignard and Cormier, 2008). To be complete, notice that in Al-rich silicate glasses (Cormier et al., 2000) and in liquid aluminates (Hennet et al., 2007, 2008; Cristiglio et al., 2008, Mei et al., 2008) the O-O distance increases up to 2.90 A due to the predominance of $AlO_4$ species in the melt.

In contrast, the oxygen coordination number is strongly composition dependent and as shown elsewhere (see Fig.5 in Guillot and Sator, 2007a) it increases almost linearly with the mole fraction of network modifiers. In fact it is the nature of the oxygen, bridging (BO) or non-bridging (NBO), and their respective population which governs the oxygen coordination number. The population of BOs and NBOs is given in Table 4 as function of composition. An oxygen atom is a BO when it is connected to two network former cations (Si or Al) whereas it is a NBO when it is connected to only one Si or Al. In our calculations two atoms (e.g. Si and O) are connected if their separation is less than the distance associated with the first minimum of the corresponding PDF. In practice this distance is 2.30A for Si-O and 2.55A for Al-O in the two simulation methods (AIMD or EPMD). For the NBOs in Table 4 we have distinguished the oxygens linked to one network former cation from the free oxygens only linked to network modifier cations. For the BOs we have distinguished the Si-O-Si, Si-O-Al, and Al-O-Al bonds as also as the triclusters O-(T)$_3$ (with T=Si or Al) which are oxygen atoms shared by three tetracoordinated cations. Although they are difficult to detect due to their very low abundance (Allwardt et al., 2003; Neuville et al., 2004), their existence is now well established, especially in Al- rich glasses (Daniel et al., 1996; Stebbins et al; 2001; Schmücker and Schneider, 2002; Iuga et al., 2005), and their NMR signature investigated by first principles calculations (Profeta et al., 2004; Benoit et al., 2005).

In silica, as expected, all oxygen atoms are BO with a very small proportion of triclusters (0.5% with AIMD and 0.1% with EPMD). In the rhyolitic melt although the alkali cations act as charge compensators for Al there is a non negligible population of NBOs (2.5% with AIMD and 4.3% with EPMD). These findings are in agreement with NMR data on charge-balanced aluminosilicate glasses (Stebbins and Xu, 1997; Stebbins et al., 2001; Allwardt et al., 2003; Neuville et al., 2004; see also in Chapter 17 of Mysen and Richet, 2005). Moreover, a small but significant population of oxygen triclusters is predicted (1% with AIMD and 1.7% with EPMD), Al and Si being about equally shared between these triclusters (see Table 4). In the basaltic liquid (An$_{36}$Di$_{64}$), the degree of depolymerization of the melt is indicated by the high abundance of NBOs (39.1% with AIMD and 38.5% with EPMD), while the population of oxygen triclusters is roughly the same than in rhyolite (~0.8-1.2%). Interestingly the NBO content is nearly that expected (NBO/$O_{Tot}$=40%) from the nominal composition and a standard model of glass structure (for a standard definition of NBO see in Chapter 4 of Mysen and Richet, 2005).

In molten enstatite, the proportion of NBOs is high (65.9% with AIMD and 61.2% with EPMD), as expected for this depolymerized melt as the standard model of glass structure predicts a value of 66.6% for the ratio NBO/$O_{Tot}$. An analysis of the atomic configurations generated by AIMD and EPMD shows that NBOs are corner- or edge-shared between $SiO_4$ tetrahedra and $MgO_{X=4,5,6}$ polyhedra. Incidentally, the amount of free oxygens (oxygen atoms only sharing Mg cations) is low, especially with AIMD (0.4%), which suggests that there is no (or very little) micro aggregation of $MgO_{X=4,5,6}$ polyhedra in the melt. Concerning the BOs, approximately 1/3 of them are linking two $SiO_4$ units exclusively, while the remaining 2/3 are shared by two $SiO_4$ units and 1- or 2- $MgO_{n=4,5,6}$ polyhedra.

From a more general point of view, further scrutiny shows that there are some subtle structural differences between the melts generated by AIMD and EPMD calculations. Some differences are visible on the PDF's (see Figs.2 and 3) but they are more conspicuous on the bond angle distributions (BAD) evaluated from the atomic configurations. In Fig.10 the BAD for Si-O-Si, Si-O-Al, and Si-O-Mg in the four simulated melts are reported. It is significant that the Si-O-Si BAD produced by AIMD is shifted toward smaller angles with respect to the distribution generated by EPMD, for all the melt compositions. This deviation between classical and ab initio calculations have also been reported in comparative simulation studies dealing with sodium tetrasilicate glass (Ispas et al., 2001, 2002) and a calcium aluminosilicate glass (Ganster et al., 2007).

In the case of vitreous silica, the archetypal network glass, the Si-O-Si BAD has been extensively studied in the literature. Although a full consensus has not been reached yet, it is admitted that the maximum (most probable value) of the BAD lies in the range 146-151° (Mozzi and Warren, 1969; Dupree and Pettifer, 1984; Pettifer et al., 1988; Poulsen et al., 1995; Neuefeind and Liss, 1996; Mauri et al., 2000; Yuan and Cormack, 2003; Clark et al., 2004; Tucker et al., 2005). However there is no consensus about the width of the distribution. In using simple models for the short range order in silica to interpret X-ray diffraction data, Poulsen et al. (1995) estimate the full width at half maximum (FWHM) about 35° when Neuefeind and Liss (1996) in using a similar approach but with combined X-ray and neutron diffraction data get a much narrower distribution with a FWHM equal to 17°. In analyzing neutron diffraction data with reverse Monte Carlo modeling, Tucker et al. (2005) extract a BAD peaking at 151° with a FWHM of 22°. On the other hand, Dupree and Pettifer (1984) and Pettifer et al. (1988) deduced the BAD from the $^{29}$Si NMR chemical shift and found a rather large variation of the distribution (a maximum between 142 and 151° and a FWHM in the range 20-25°) with the model used to relate the Si-O-Si bond angle and the chemical shift. More recently, progress in $^{17}$O NMR spectroscopy permit to obtain 2D histograms of bridging oxygen structural parameters (Clark et al., 2004). The deduced BAD peaks at 147° but is significantly narrower (~10°) than found in other studies of the literature.

The results with AIMD leads to a value around 132° for the most probable Si-O-Si bond angle in liquid silica (in agreement with previous ab initio studies by Sarnthein et al., 1995, and Karki et al., 2007), whereas the EPMD simulation gives ~147°, a value within the experimental interval, and which is the value generally obtained with commonly used 2-body potentials for silica (e.g. Tsuneyuki et al., 1988; van Beest et al., 1990). In fact the reduction of the most probable bond angle is the direct consequence of a slightly larger value of the [4]Si-O bond distance in the ab initio calculation with respect to the classical one (1.69 A instead of 1.67 A) in conjunction with a slightly smaller Si-Si distance (3.15 A instead of 3.17 A, see Fig.2). As a matter of fact, the maximum of the Si-O-Si BAD is governed by the intra tetrahedral Si-O bond distance and the inter tetrahedral Si-Si distance (O'Keeffe and Hyde, 1978). With regard to the width of the distribution, it is associated with Si-2nd O, Si-2nd Si and O-2$^{nd}$ O inter tetrahedral distances (Yuan and Cormack, 2003). The widths generated by our AIMD and EPMD simulations in liquid silica are broad and quasi identical to each other with a FWHM about 43°. Although one may expect that the BAD broadens somewhat between the vitreous state and the high temperature liquid, our conclusion is that the present simulations overestimate the width in liquid silica. As a matter of fact, we have mentioned in the previous section, that the third peak of the X-ray RDF is localized at a higher r value in the calculations with respect to the experimental feature (4.3 A instead of 4.15 A, see Fig.4). Since the third peak is produced by the overlap of the inter tetrahedral Si-2$^{nd}$ O, Si-2$^{nd}$ Si and O-2$^{nd}$ O distances, the deviation in position (by +0.15 A) is a strong indication that the atomic configurations show a too large distribution of inter tetrahedral angles with respect to those probed by diffraction data.

The presence of a few percent of alkali ions in rhyolite changes markedly the Si-O-Si BAD when comparing with that exhibited by silica. Thus the maximum of the distribution shifts from 132° to 124° with AIMD and from 147° to 140° with EPMD (see Fig.10). However the shape of the BAD becomes skewed and narrower (FWHM~33°) with AIMD when it is practically invariant with EPMD (FWHM~43°). Correspondingly, the Si-Si distance evaluated by AIMD calculation decreases from 3.15A in silica to 3.05 A in rhyolite, whereas this distance is practically invariant with EPMD (see Fig.2). In fact, in decreasing the strain along the tetrahedral chains, the NBOs are responsible of the low angle shift of the BAD. From the experimental viewpoint, the NMR study of $K_2SiO_4$ glass by Farnan et al. (1992) shows a Si-O-Si angle distribution which is skewed and shifted toward a lower angle as compared with silica. In the same way, Zotov and Keppler (1998) by investigating a sodium tetrasilicate glass by neutron diffraction experiment, extract from their data via a reverse Monte Carlo calculation, a Si-O-Si BAD very much alike the one obtained by AIMD, that is exhibiting an asymmetric shape with a maximum near 125°. Another piece of information is given by the simulation study of Yuan and Cormack (2003) where it is pointed out the relationship between the changes of the BAD on adding sodium and the rising of a hump around 3.7 A in the pair correlation function of alkali silicate glasses investigated by neutron scattering (Wright et al., 1991). In our AIMD calculation, a shoulder is quite visible near 3.5 A in the Si-O pair distribution function (see Fig.2), the dominant contribution to the neutron correlation function in this R- range. This Si-O distance is characteristic of a staggered configuration between oxygens of two $SiO_4$ tetrahedra sharing the same BO. By contrast, the EPMD calculation produces no shoulder on the Si-O PDF. These findings support the statement of Yuan and Cormack (2003) that a pronounced shoulder on the Si-O PDF indicates a narrower BAD (e.g. AIMD) and a smooth shoulder requires a broader BAD (e.g. EPMD). In summary, the structure of liquid rhyolite generated by AIMD is much more sensitive to the presence of alkali ions than the one generated by EPMD in spite of its smaller proportion of NBOs (2.5% instead of 4.3%). The success of the AIMD calculation to reproduce the evolution of the melt structure on adding alkali ions expresses its ability to cope with the ionicity-covalency ambivalence of the interactions in silicate melts.

Curiously enough, the Si-O-Si BAD, as evaluated by AIMD, changes very little in going from the felsic composition to the basic and ultrabasic one (see Fig.10) whereas with EPMD a slight shift of the distribution toward smaller angles is observed. This finding suggests that the Si-O-Si BAD is a complex function of the amount of NBOs in the melt and of the nature of the cations. In the case of Al, the Si-O-Al BAD evaluated by AIMD is more dependent on composition than the one evaluated by EPMD (see Fig.10). Furthermore, the maximum of the distribution is localized at a lower angle than the one exhibited by the Si-O-Si BAD (in the range 120-130° instead of 124-140°). This shift is correlated with the length of the Al-O bond which is larger than the Si-O one while Si-Si and Si-Al distances are very close to each other and virtually invariant with composition (see Figs.2 and 3). Moreover the maximum of the Si-O-Al BAD tends to shift to a smaller angle from felsic to basic liquid. These findings are in a qualitative agreement with the study of Taylor and Brown (1979b) on feldspar glasses who observed that the T-O-T (T=Si, Al) bond angle inferred from X-ray diffraction data decreases from alkali to calcic glasses.

In molten enstatite, a majority of oxygen atoms connects a $SiO_4$ tetrahedron to $MgO_{X=4,5,6}$ polyhedra. The Si-O-Mg BAD expresses this connectivity through a broad distribution which is essentially featureless except two faint maxima at 90° and 120°. The maximum at 120° can be attributed to corner sharing $SiO_4.MgO_4$ configurations whereas the maximum at 90° correspond to edge sharing $SiO_4.MgO_{4,5,6}$ configurations.

### 3.3 Self-diffusion coefficients

Because ionic diffusion in silicate melts plays a key role in a number of geological processes (e.g. rheology of magmas, the extent to which equilibrium may have been reached during crystallization and melting, etc.) it is important to evaluate how is the accuracy of the AIMD and EPMD simulations to reproduce the ionic diffusivity data. Although there exists an abundant literature on diffusion coefficients of various ionic species in synthetic or natural silicate melts (for reviews see Brady, 1995; Zhabrev and Sviridov, 2003), mostly, the diffusivities of the major cations are not measured altogether for a given melt. Furthermore, the temperature of the investigated melt rarely exceeds 1500-1600°C in these studies (for exceptions see Poe et al., 1997; Reid et al., 2001). On the other hand, for reason which will become clearer later on, MD simulations are generally performed at very high temperatures well above 2000°C (e.g. Angell et al, 1982; Bryce et al., 1997; Lacks et al., 2007), which makes the comparison with experimental data obtained at a much lower temperature in the supercooled liquid or in the glass difficult.

We have evaluated the self-diffusion coefficients of all ionic species present in the simulated melt from the relation (Kubo, 1966),

$$D_s = \lim_{t \to \infty} \frac{1}{Ns} \sum_{i=0}^{Ns} \frac{<(ri(t) - ri(0))^2>}{6t} \qquad (1)$$

where $r_i$ is the position of the ion $i$ of species $s$, and where the bracket expresses an ensemble average taken over many origin times. For EPMD simulations, long runs have been performed (from 10ns for rhyolite, $An_{36}Di_{64}$, and enstatite, up to 100ns for silica) to be sure that the diffusive regime was reached (in this regime the mean square displacement increases linearly with time, which allows one to evaluate $D_s$ from eqn.(1)). For AIMD simulations, the cost in computer time is much too high to investigate the nanosecond time scale, and only run lengths of ~5 ps have been performed. This limitation in computer time has important implications because for a very viscous liquid such as silica at 2273K ($\eta$=6.1 $10^4$ PaS after Urbain et al., 1982), our EPMD simulations indicate that the diffusive regime for the displacement of Si and O atoms is reached only after 10ns, and solely the rattling motions of Si and O atoms (caging effect) can be investigated by AIMD simulation. This is illustrated in Fig.11 where the log-log plot of the mean square displacement (MSD) versus time is shown, the log-log plot being the best representation to detect the diffusive regime as the MSD increases linearly with time in that case.

From a numerical viewpoint it is noteworthy that finite size effects due to the limited number of simulated atoms are non negligible when investigating diffusion coefficients by molecular dynamics simulation (Yeh and Hummer, 2004). Thus for liquid silica, it is known that significant finite size effects are present in the evaluation of the self diffusion coefficient when dealing with system sizes much smaller than 1000 atoms (Horbach et al., 1996; Saksaengwijit and Heuer, 2007). Recently it has been established for liquid silica simulated with the BKS potential (Zhang et al., 2004) that the logarithm of the diffusion coefficients for Si and O atoms scale linearly with $1/N^{1/3}$, where N is the number of simulated atoms, at very high temperature in a region where liquid silica exhibits a "fragile behavior" (see Saika-Voivod et al., 2001) and with 1/N at low temperature in a region where liquid silica exhibits a "strong behavior" Arrhenius-like. Thus, according to the results of Zhang et al. (2004) our EPMD simulations with N~100 atoms underestimate the diffusion coefficients of Si and O in liquid silica at 2273K by a factor of ~3-4. We have checked this point for rhyolite, $An_{36}Di_{64}$ and enstatite at 2273K by comparing the results of EPMD simulations for two system sizes with 100 and 1000 atoms, respectively. The size effects are found to be strongly dependent on the viscosity of the melt under investigation; at a given temperature they decrease from felsic to ultrabasic composition. Moreover, the size effects are larger for network former ions than for network modifiers. Thus, in rhyolite, the ratio of the diffusion coefficients $D_s^{N=1000}/D_s^{N=100}$ is equal to ~3 for Si, O and Al atoms, but only ~1.6 for K and ~1.2 for Na. In the anorthite-diopside melt this ratio amounts to ~1.2 for Si, O and Al atoms whereas there are no detectable size effects with Mg and Ca atoms. In the ultrabasic melt the self diffusion coefficients for Si, O and Mg atoms are virtually independent of the size of the simulated system. Although we are unable, at the present time, to evaluate the size effects for AIMD simulations, we expect that the trends observed with EPMD might occur with AIMD. In the following, it will be important to keep this information in mind.

Experimentally, Mikkelsen (1984) reported values of the self diffusion coefficient of network oxygen in vitreous silica between 1200 and 1400°C. A simple linear extrapolation up to 2273K in using its Arrhenius plot for D versus 1/T leads to an oxygen diffusivity about $1.\times10^{-14}$ $m^2/s$ as compared with $130.\times10^{-14}$ $m^2/s$ obtained by EPMD simulation. As for the diffusivity of Si, we obtain $71\times10^{-14}$ $m^2/s$ when an Arrhenian extrapolation from the data by Brebec et al. (1980) and Takahashi et al. (2003) gives $D_{Si}$=0.15 and $0.02\times10^{-14}$ $m^2/s$, respectively. However, the experimental values of $D_{Si}$ and $D_O$ published in the literature show a considerable variation between studies (e.g. Kalen et al., 1991; Mathiot et al., 2003; Fukatsu et al., 2003; Uematsu et al., 2004), and some care must be taken when doing high temperature extrapolation. An independent evaluation of the order of magnitude of $D_{Si}$ and $D_O$ in liquid silica at high temperature can be deduced from the empirical model of Mungall (2002) relating viscosity and tracer diffusion for a large variety of magmatic silicate melts. Although the model predicts the same diffusivity for all network former ions ($D_{Si}\sim D_O\sim D_{Al}$) it fits remarkably well a large set of diffusivity data for ions in silicate melts of rhyolitic to basaltic composition. Thus, in using the viscosity data of Urbain et al. (1982) for liquid silica (for a critical discussion of viscosity data, see Doremus, 2002), the model of Mungall (2002) predicts a diffusivity for Si and O about $0.61\times10^{-14}$ $m^2/s$ at 2232K, a value which compares well with the data of Mikkelsen (1984) and Brebec et al. (1980). In summary, we conclude that our simulated silica (EPMD) is not sufficiently viscous because the calculated diffusivities for Si and O atoms are overestimated by two or three orders of magnitude at ~2273K (all data are collected in Table 5). This inability of classical simulations using empirical potentials to model highly viscous liquids is well documented in the literature (for a discussion see Micoulaut et al., 2006). Unfortunately, as mentioned earlier, the hope to check whether the AIMD simulation can remedy to this situation is still out of reach.

For rhyolite, the situation is different because the deviation between calculated diffusivities and those measured in melts of rhyolitic composition depends on the investigated elements (see Table 5). Here again, the comparison is based on experimental values measured either in the vitreous state or in the supercooled liquid and extrapolated at high temperature to about 2273K. Hence one may conclude that the EPMD simulation overestimates the diffusivity of Si, O and Al atoms at this temperature by more than one order of magnitude (including the finite size effects). For Na and K the agreement is rather satisfactory (taking into account all sources of uncertainties). Moreover, the simulation shows that the mobility of Na and K atoms are essentially decoupled from that of network former ions, a feature well known from experimental data (Dingwell and Webb, 1990; Gruener et al., 2001; Meyer et al., 2002; Gaillard 2004). For instance, in increasing the temperature of the simulated melt by 200K, the Si, O and Al diffusivities increase by roughly a factor of 3 while those of Na and K barely increase (~10%). Furthermore, a comparative analysis of the time evolution of the MSD for Na and network former ions shows that Na ion dynamics relaxes on a time scale of 10ps when the network relaxes on a time scale of ns (see Fig.11), Na ions diffusing in a quasi immobile matrix. This result is in agreement with the conclusion of an inelastic neutron scattering study (Meyer et al., 2002) of a sodium disilicate melt and supports the picture where the characteristic time for sodium hopping is much shorter than the structural relaxation time (Gruener et al., 2001). With regard to our AIMD calculations, it is not possible to observe the diffusion of atoms in the limited computer time. However, because rhyolite has much lower viscosity than silica (e.g. $\eta$=10 PaS at 2473K instead of ~4400 PaS for silica), it is possible to reach by AIMD simulation the pre-diffusive regime occurring in the ps time scale: this is clearly seen in Fig.11. In this time domain, the MSD for Si, O and Al atoms are very near those evaluated from the EPMD simulation, whereas those of Na and K are lower with AIMD.

For molten $An_{36}Di_{64}$, the diffusivities of Si and O evaluated by EPMD (see Table 5) are larger, by may be a factor of 3~5 (including the finite size effects), than those measured for an anorthite-diopside melt of related composition ($An_{42}Di_{58}$ by Dun, 1982; Tinker et al., 2003), or evaluated by the diffusivity-viscosity relationship of Mungall (2002) in using the viscosity model of Russel and Giordano (2005). Thus the model of Mungall (2002) predicts a value $D_{Al}$~$2D_{Si,O}$=$0.9x10^{-9}$ $m^2$/s in $An_{36}Di_{64}$ (for a viscosity $\eta$=0.19 PaS at 2273K) whereas we obtain $D_{Al}$~$2D_{Si}$=$1.9x10^{-9}$ $m^2$/s. In the same way, the calculated diffusivities for Mg and Ca are about three times those measured in a haplobasaltic melt by La Tourette et al. (1996). The AIMD simulation leads to a real improvement since the examination of the MSD in the pre-diffusive regime (see Fig.11) shows that the diffusivities are significantly reduced with the quantum calculation, may be by a factor of ~2-3 for Si, O and Al and by a factor of ~5 for Mg and Ca. Hence one may conclude that the rheological properties of the quantum simulated anorthite-diopside melt might be realistic.

To our knowledge there is no diffusivity data available in the literature for molten enstatite. Therefore, we have used for comparison the diffusivity data of Si, O and Mg in molten diopside (Dunn, 1982; Shimizu and Kushiro, 1991; Reid et al., 2001), a basic melt whose the viscosity is very close to that of molten enstatite (at 2200K, $\eta$(enstatite)=$8.x10^{-2}$ PaS and $\eta$(diopside)=$7.x10^{-2}$ PaS after Urbain et al., 1982). As reported in Table 5, the Si and O diffusivities calculated by EPMD simulation at 2573K ($1.8x10^{-9}$ $m^2$/s and $2.6x10^{-9}$ $m^2$/s, respectively) are in agreement both with the diffusivities measured by Reid et al. (2001) in molten diopside (~$5.2x10^{-9}$ $m^2$/s, after an Arrhenian extrapolation between 2273 and 2573K), and those deduced from the model of Mungall ($D_{Si,O} \sim 2.7 \times 10^{-9}$ m$^2$/s). A good agreement is also found for the self diffusion of Mg because the value calculated at 2573K ($\sim 8.7 \times 10^{-9}$ m$^2$/s) is near the estimation made by Shimizu and Kushiro (1991) in molten diopside ($\sim 5.5 \times 10^{-9}$ m$^2$/s after temperature extrapolation). As for the AIMD simulation, the analysis of the pre-diffusive regime exhibited by the MSD of Si, O and Mg atoms in the quantum simulated melt indicates lower values of the diffusivities as compared with those observed with the classical simulation (see Fig.11). Tentatively, we propose the following values at 2573K: $D_{Si} \sim 1. \times 10^{-9}$ m$^2$/s, $D_O \sim 1.5 \times 10^{-9}$ m$^2$/s and $D_{Mg} \sim 3 \times 10^{-9}$ m$^2$/s. However, the lack of ionic diffusivity data in molten enstatite precludes to give a more definite conclusion about which calculation (EPMD versus AIMD) is the most realistic.

### 3.4 Infrared absorption spectrum

Vibrational motions of atoms in silicate melts and glasses have been investigated for over fifty years by infrared and Raman spectroscopies (for reviews see Mysen et al., 1982; Mc Millan and Wolf, 1995; King et al., 2004) which probe the frequency domain 5-5000 cm$^{-1}$ (or the ps-fs time scale). In infrared spectroscopy (IR), the ionic motions are revealed through the time fluctuations of the total dipole moment of the sample resulting from the charge distribution in the melt, whereas in Raman spectroscopy the distortions of the ionic polarizabilities driven by the atomic displacements are responsible for the spectrum (Fowler and Madden, 1985). Because the evaluation of the Raman spectrum by MD simulation is a difficult task involving significant approximations, the calculations presented below concern only the evaluation of the IR absorption spectrum.

In statistical mechanics the IR absorption coefficient is given by (McQuarrie, 1976),

$$\alpha(\omega) = \frac{\omega Im. \varepsilon(\omega)}{n(\omega)c} = \frac{2\pi\omega}{3\hbar n(\omega)cV}\left(1 - e^{-\hbar\omega/kT}\right)\int_{-\infty}^{+\infty} e^{-i\omega t} < M(t).M(0) > dt \qquad (2)$$

where $Im.\varepsilon(\omega)$ is the imaginary part of the frequency dependent dielectric constant, $n(\omega)$ the frequency dependent refractive index, $\hbar$ the Planck constant divided by $2\pi$, $k$ the Boltzmann constant, $c$ the speed of light and $V$ the volume of the sample. In eqn.(2), $M(t)$ is the total dipole moment of the liquid sample and the bracket expresses a canonical average.

In EPMD simulation, the ions are assumed to be rigid (nonpolarizable) with an effective charge $z_i$. Hence, the expression of the dipole moment of the simulated sample at time t is,

$$M(t) = \sum_i z_i \, r_i(t) \tag{3}$$

where the sum runs over all the ions of the sample and where $r_i(t)$ is the position of the ion $i$ at the simulation step t. To improve the statistics it is convenient to introduce the total ionic current,

$$J(t) = \frac{dM(t)}{dt} = \sum_i z_i \, v_i(t) \tag{4}$$

where $v_i(t)$ is the velocity of the ion $i$. In order to substitute in eqn.(2), M(t) for J(t), it is noteworthy that ,

$$I_M(\omega) = I_J(\omega)/\omega^2 \tag{5}$$

where $I_M(\omega)$ and $I_J(\omega)$ are the Fourier transforms of the time autocorrelation functions associated with M(t) and J(t), respectively. In the limit where quantum effects can be neglected (this is the case for the high temperature liquid in the frequency range 0-1500cm$^{-1}$ investigated here, where $\hbar\omega/kT \ll 1$ ), the introduction of eqn.(5) into eqn.(2) gives the simple relation,

$$\alpha(\omega) = \frac{4\pi^2}{3kTn(\omega)cV} I_J(\omega). \tag{6}$$

In AIMD simulation, the total dipole moment M is the sum of an ionic contribution (because of the distinction made between core electrons and valence electrons, each atom has an ionic core) and an electronic contribution. When the calculation of the IR absorption coming from the ionic part of the dipole moment is analog to the procedure described above for EPMD, the evaluation of the electronic contribution is not straightforward. In fact the electronic dipole cannot be defined unambiguously for an infinitely replicated system as the one used here. This problem has been solved in using the Berry phase scheme (Vanderbilt and King-Smith, 1993) and the IR spectra of liquid water, amorphous silicon, and silica glass have been successfully evaluated by AIMD in this framework (Silvestrelli et al., 1997; Debernardi et al., 1997; Pasquarello and Car, 1997). To avoid a long digression, the reader is referred to these works for technical details. Moreover, the short runs performed with AIMD simulations generate significant computational noise and, to improve the statistics, it is convenient to evaluate numerically the time derivative of the electronic polarization. Therefore, the absorption coefficient is evaluated from eqn.(6) where $I_J(\omega)$ is the Fourier transform of the generalized current,

$$J(t) = \frac{d}{dt}\{M_{ion}(t) + M_{elec}(t)\} = J_{ion}(t) + J_{elec}(t) \qquad (7)$$

where $J_{ion}(t) = \sum z_i v_i(t)$ is calculated from the velocities of the ionic cores bearing a charge $z_i$, and where $J_{elec}(t)$ is evaluated numerically from $M_{elec}(t)$ by means of finite difference between time step (t) and time step (t -1).

In practice, the autocorrelation function for $J(t)$ was evaluated along simulation runs of 50 ps by EPMD calculations whereas the length of the runs performed with AIMD were only 5ps. Even if these runs are short this is sufficient to evaluate the IR spectrum with a reasonable accuracy (±10%) provided that $J(t)$ is evaluated for different time origins along the MD run. Indeed, most of the IR spectral intensity measured with silicate melts is in the frequency range 5-1500 $cm^{-1}$, which corresponds to the time domain 3fs - 1ps. The results of our calculations for the four compositions are discussed below.

**3.4.1 Silica**

Silica glass is prototypical of silicate glasses and for this reason its IR spectrum is used as a benchmark. It is dominated by three main bands (see in Fig.12 the absorption spectrum by Velde and Couty, 1987): an intense high frequency band peaking about 1100 cm$^{-1}$ with a shoulder near 1200 cm$^{-1}$, weakly IR active, caused by the stretching modes of $SiO_4$ units, a weak band at intermediate frequency (~800 cm$^{-1}$) expressing the vibration motions of Si in their tetrahedral cage and an intense low frequency band (~470 cm$^{-1}$) resulting from bending modes of O-Si-O bonds. Other weak bands and shoulders can also been identified around 400 and 600 cm$^{-1}$ and an abundant literature is devoted to the interpretation of all these bands in terms of structural units (Gaskell and Johnson, 1976; Mysen et al., 1982; Kirk, 1988; McMillan and Wolf, 1995; Sarnthein et al., 1997; Dalby et al., 2006). Upon heating, Markin and Sobolev (1960) observed a significant low frequency shift of the high frequency band from 1120 cm$^{-1}$ in the glass at room temperature to 1054 cm$^{-1}$ in the liquid at 2273K and a decrease of its intensity by roughly a factor of two. For the other bands we have no information about the effect of the temperature on the IR spectrum even if a broadening of bands is expected from glass to liquid silica. Nevertheless, a Raman study between room temperature and 1950K (McMillan et al., 1994) shows a slight high frequency shift of the band at 450 cm$^{-1}$ upon heating and a quasi invariance of the intermediate frequency band with temperature whereas the high frequency band becomes too weak to be clearly identified above ~1800K. Notice that the lack of IR data at high temperature is the current situation for silicates and hence in the following we will compare our results with IR data obtained with glasses.

The IR spectrum of liquid silica evaluated by EPMD simulation shows essentially two bands (see Fig.12), one peaking around ~1000 cm$^{-1}$ and the other one around ~550 cm$^{-1}$. In order to estimate the role of the temperature on the spectral profile, we have also evaluated the absorption coefficient of the glass by cooling the liquid down to 300K with a quenching rate as low as 10$^{10}$ K/s (in general much higher quenching rates, ~10$^{14}$ K/s, are used in the simulation literature, see for example Hemmati and Angell, 1997). From liquid to glass, one notices (see Fig.12) an enhancement of the high frequency band which is shifted from ~1000 to ~1100 cm$^{-1}$ with the occurrence of a shoulder near 1260 cm$^{-1}$ when the low frequency band decreases somewhat in intensity and exhibits a shift from ~550 to ~520 cm$^{-1}$. In addition, an intermediate band shows up around ~760 cm$^{-1}$. Although the simulated spectrum shares some similarities with the experimental one it does not fit well (see Fig.12). The low and intermediate frequency bands are too broad and the intensity of the high frequency stretching band is much too weak. These kind of discrepancies are known for rigid ion models where polarization effects are not explicitly taken into account (see Wilson et al., 1996; Hemmati and Angell, 1997; Guillot and Guissani, 1997).

In contrast, the IR spectrum of the liquid phase evaluated by AIMD simulation presents two intense frequency bands at low and high frequency with a shoulder on the low frequency flank of the Si-O stretching band. Moreover, the intensity of the latter one has the accurate order of magnitude. Indeed, experimentally the absorption coefficient in vitreous silica reaches ~40,000 cm$^{-1}$ at the maximum of the high frequency band (Gervais et al., 1987). Knowing that the absorption intensity of the high frequency band decreases by a factor of ~2 from the glass to the liquid at 2000°C (Markin and Sobolev, 1960), the absorption intensity of the high frequency band is expected to be of the order of ~20,000 cm$^{-1}$ for the liquid, virtually the same value as obtained by AIMD simulation (~20,500 cm$^{-1}$, see Fig.12). However, the three main bands of the calculated spectrum are not located at the correct frequencies (~400, 700 and 870 cm$^{-1}$, instead of ~470, 800 and 1050 cm$^{-1}$ in the vitreous state). This shortcoming of the ab initio calculation is puzzling because Pasquarello and Car (1997) have obtained by AIMD simulation a better agreement for the position of the frequency bands in the IR spectrum of amorphous silica. In fact it is the use of the generalized gradient approximation (GGA) in our electronic structure calculation which is the origin of the disagreement about the vibration frequencies. As mentioned earlier (see section 2) GGA tends to underestimate the binding of the silica network when LDA, as used by Pasquarello and Car, tends to overestimate it (see also Karki et al., 2007). Thus the Si-O distance is evaluated about 1.65A with GGA when LDA leads to a value very close to the expected one (~1.62A, Tucker et al. 2000), an overestimation which lowers the vibration modes associated with $SiO_4$ units in GGA calculation.

In order to quantify the contribution of the electronic polarization to the IR spectrum, we have evaluated separately the absorption coming from the ionic cores (ionic contribution). For a better comparison with the EPMD calculation where the ions are considered as non polarizable, we have evaluated the ionic contribution to the IR spectrum in using the atomic configurations generated by the AIMD simulation and by assigning to the Si and O ionic cores the effective charges used in the EPMD simulation (i.e. $q_{Si}$=1.89e and $q_O$=-0.945e). The resulting spectrum shown in Fig.12 (labeled AIMD (no pol.) in the figure) is quite different from the AIMD spectrum evaluated with the electronic polarization. In particular, the intensity of the bands decreases drastically when the polarization is not accounted for but their position in frequency is not modified. A further analysis shows that when the electronic polarization is included some vibrations become more infrared active than others: this is the case of the Si-O stretching mode at high frequency. This propensity of the electronic polarization to redistribute the spectral intensity is a key feature of interaction induced spectra encountered with a large variety of molecular systems (for reviews about interaction induced spectra in molecular liquids see Tabisz and Neumann, 1995). In conclusion, as long as the IR spectrum is concerned, the AIMD simulation, by taking into account the electronic polarization from the first principles, is clearly a major improvement to classical simulations using the rigid ion approximation.

### 3.4.2 Rhyolite

In the experimental literature, the IR spectra of rhyolitic glasses (e.g. Houziaux, 1956; Shimoda et al., 2004; Byrnes et al., 2007) are similar to that reported for silica glass except that the two main vibration bands, at about ~470 cm$^{-1}$ and ~1070 cm$^{-1}$, are less intense (may be by a factor of ~2, according to a comparison made by Gervais et al. (1987), between a sodium silica glass and vitreous silica, see Fig.10 in the aforementioned article) and are broader for rhyolite than for silica (compare the experimental spectra shown in Figs.12 and 13). Moreover the high frequency band in rhyolitic glasses (e.g. fused obsidian) is shifted by about -50 cm$^{-1}$ with respect to that observed in vitreous silica (1070 cm$^{-1}$ versus 1120 cm$^{-1}$) and presents a pronounced shoulder near 1200 cm$^{-1}$. At low frequencies in the far infrared range (0~300 cm$^{-1}$), the absorption intensity is expected to be much more effective in glasses of rhyolitic compositions than in silica glass, as it is observed with alkali pentasilicate glasses (Kamitsos and Risen, 1984) and with Na-bearing aluminosilicate glasses (Gervais et al., 1987; Merzbacher and White, 1988).

As long as the absolute value of the absorption intensity is concerned, that calculated by AIMD for liquid rhyolite is comparable to IR intensities measured in sodium silica glasses whereas that calculated by EPMD is much too weak. As a matter of fact, Gervais et al. (1987, see Fig.3 therein) have measured an absorption intensity about 22,500 cm$^{-1}$ at the maximum of the high frequency band of a sodium silica glass whereas the intensity of this band amounts to ~16,500 cm$^{-1}$ in liquid rhyolite by AIMD simulation and only ~2,000 cm$^{-1}$ by EPMD simulation (see Fig.13). As in the case of silica, the contribution of the electronic polarization to the absorption intensity is predominant. Without electronic polarization, the calculated intensity decreases drastically and becomes as low as the one obtained by EPMD (see Fig.13), the drop of the intensity being much larger for the high frequency band than for the low frequency one (a factor of ~5 instead of a factor of ~2). Nevertheless, as in the case of silica, the location in frequency of the IR absorption bands predicted by the AIMD simulation does not coincide accurately with the observed bands (e.g. the Si-O stretching band peaks at 830 cm$^{-1}$ instead of 1070 cm$^{-1}$ with a shoulder near 1100 cm$^{-1}$ instead of 1200 cm$^{-1}$).

To understand better the genesis of the band shape we have evaluated the partial spectra associated with the different ionic species present in the simulated melt (AIMD and EPMD). These spectra were evaluated by taking the Fourier transform of the autocorrelation functions of the partial ionic currents, $J_s(t) = \sum_i z_i\ v_i(t)$, where $i$ runs over all ions of species $s$. Note that this calculation of the partial ionic spectra is just a guide to identify rapidly the ionic motions at the origin of the band shape, and a detailed account of the various vibration modes associated with ionic species and structural units present in the melt is beyond the scope of this article. As illustrated in Fig.14, the motions of Si and O atoms are the main contributors to the absorption spectrum of rhyolite over the whole frequency range. As for the motions of Al, they are characterized by a partial spectrum, broad and asymmetric, quite different from those of Si and O. The corresponding absorption intensity is mainly below $800 cm^{-1}$, which is expected as Al is less tightly bound to oxygen atoms than Si. Furthermore the vibration motions of $AlO_{4.5}$ units in the melt simulated by AIMD are characterized by a stretching mode around $600\ cm^{-1}$ and bending modes around 300 and $200\ cm^{-1}$, at variance with the EPMD calculations for which no clear excitation occurs in these frequency domains (see Fig.14). Notice that Gervais et al. (1987) have investigated a large number of alkali aluminosilicate glasses containing between 3 and 43% of aluminum and no new IR band characteristic of this element has been observed although a global increase of the absorption intensity in the intermediate frequency range is conspicuous. At lower frequencies in the far infrared range ($\omega < 300\ cm^{-1}$), the simulations predict a significant absorption intensity due to the vibration modes of Na and K peaking near $100 cm^{-1}$. The latter modes have been observed in alkali aluminosilicate glasses (Gervais et al., 1987; Merzbacher and White, 1988) and in germanate glasses (Kamitsos et al., 1996).

### 3.4.3 Anorthite-diopside eutectic

The anorthite-diopside eutectic ($An_{36}Di_{64}$) sometimes is considered as a model basalt (e.g. Rigden et al., 1984). The IR absorption spectrum of the glass is characterized by a very broad high frequency band (FWHM ~ 400 $cm^{-1}$ as compared with ~ 160 $cm^{-1}$ for silica glass) showing a round maximum at about 1000 $cm^{-1}$ (Taniguchi and Murase, 1987). Unfortunately, at our knowledge, the absorption spectrum of the glass has not been recorded below 800 $cm^{-1}$. However, the IR spectra of diopside glass (Kubicki et al., 1992; Koike et al., 2000; Shimoda et al., 2005) and glassy basalts (Crisp et al., 1990; Minitti et al., 2002; King et al., 2004; Johnson et al., 2007) are assumed to be very similar to that produced by the anorthite-diopside glassy mixture. As a matter of fact, the IR spectra of the former ones are characterized by a very broad high frequency band (FWHM~400 $cm^{-1}$ in diopside glass and ~450 $cm^{-1}$ in basalt with $\omega_{max}$~1020 and 950 $cm^{-1}$, respectively) and a low frequency band (FWHM~200 $cm^{-1}$) centered around 500 $cm^{-1}$, with significant absorption intensity in between (see Fig.15). Nevertheless, it is noteworthy that the IR spectrum of diopside glass shows a deficit of absorption in the intermediate frequency range with respect to the spectrum of glassy basalt likely because of the presence of aluminum in the latter one. By comparison, the theoretical (AIMD) spectrum of $An_{36}Di_{64}$ is rather similar to the IR spectra described above except that the entire band shape is shifted to lower frequencies by ~-130 $cm^{-1}$. Thus the high frequency band is located around 870$cm^{-1}$ instead of ~1000 $cm^{-1}$, and the low frequency band peaks about ~370 $cm^{-1}$ instead of 500 $cm^{-1}$. Moreover, pronounced absorption intensity is obtained in the intermediate frequency range (500-700 $cm^{-1}$). In contrast, the EPMD simulation predicts a better location in frequency of the two main vibration bands (see Fig.15), but their intensity is much too weak as the electronic polarization is not accounted for in this model.

The analysis of the individual ionic currents (Fig.14) shows that the vibration modes of $AlO_{4,5}$ units contribute significantly to the absorption intensity in the frequency range 100-700 $cm^{-1}$ and particularly in the frequency region separating the low frequency band from the high frequency one. Besides, the presence of infrared bands in the 500-900 $cm^{-1}$ frequency range observed with Al- rich aluminosilicate glasses was also assigned to the vibration modes of $AlO_{4,5,6}$ polyhedra (Poe et al., 1992b). As for the vibration modes of Mg ($\omega_{max}\sim250cm^{-1}$ with AIMD) and Ca ($\omega_{max}\sim180\ cm^{-1}$ with AIMD) they are responsible of broad absorption intensity in the far infrared range. These assignments are in excellent agreement with those based on far infrared data in Ca- and Mg- bearing aluminosilicate glasses (Gervais et al., 1987).

### 3.4.4 Enstatite

Spectroscopic observations in the IR range of interstellar grains and dust particles have shown that they are composed of amorphous and crystalline Mg-rich silicates, mainly olivines and pyroxenes (Molster et al., 2002). In this context the IR absorption spectrum of enstatite glass has been measured in the laboratory by several authors (Stephens and Russell, 1979; Koike et al., 2000; Brucato et al., 2000) to be used as a reference spectrum. As shown in Fig.16, it is characterized by two main bands, an intense low frequency band ($\omega_{max}\sim520\ cm^{-1}$) with a tail extending to the far infrared range, and a broad and intense high frequency band ($\omega_{max}\sim1020\ cm^{-1}$). Between these two frequency bands in the frequency range 600-700 $cm^{-1}$ the spectrum is rather flat and structureless. The IR spectra evaluated by EPMD and AIMD simulations differ significantly from each other. While the AIMD simulation reproduces satisfactorily the order of magnitude of the absorption intensity of the infrared band, the absorption intensity generated by EPMD simulation is much too weak and especially the high frequency band (see Fig.16). The same disagreement is observed when the electronic polarization is not accounted for in the AIMD calculation. However, the position of the two main bands is not predicted with accuracy by the AIMD calculation (around 370 and 860 $cm^{-1}$ as compared with 520 and 1020 $cm^{-1}$, experimentally).

The evaluation of the partial spectra (see Fig.14) shows that $MgO_{4,5,6}$ polyhedra contribute significantly to the absorption intensity in the far infrared range (0-300 $cm^{-1}$) and in the intermediate frequency region (300-700 cm$^{-1}$) as well, a finding in agreement with IR data on magnesian silicate glasses (Gervais et al., 1987). A further scrutiny of Fig.14 indicates that three Mg-O vibration modes are particularly infrared active at ~160, 300 and 500 cm$^{-1}$ (for AIMD) but they are barely visible in the total spectrum due to the presence of Si-O bending modes in this frequency range. As for the bands localized at higher frequency, they are produced essentially by bending and stretching modes of $SiO_4$ units involved in various $Q_n$ species (with n<4), where n is the number of BOs.

## 4. Conclusion

Ab initio molecular dynamics simulations have been performed to investigate the structure and dynamical properties of four silicate liquids covering a range of composition representative of the variety of natural magmas. It is shown that the heuristicity of the quantum calculation is remarkable considering that, without any ad hoc adjustment, it is able to reproduce fairly well the evolution of the structural and dynamical properties with melt composition. However, the comparison with experimental data and simulation results obtained with an empirical force field developed specifically to best describe the silicate melts belonging to the KNFCMATS system, shows that there is a need for improvement. Thus, although the two calculations (AIMD versus EPMD) produce a microscopic structure very close to each other, there are some subtle differences. For instance, the Si-O-Si bond angle distribution generated by AIMD agrees less with the available experimental data than the EPMD calculation (the angle corresponding to the distribution maximum is too small). This deviation is correlated with Si-O and Al-O first neighbor distances which are a bit too large (by ~0.02-0.03 A) in the quantum calculation. Consequently, the vibrations associated with $TO_4$ units are characterized by frequency modes which are too low (by ~150 cm$^{-1}$) with respect to those identified in IR spectroscopy data. These inaccuracies of the ab initio method need to be corrected in the future. Thus, it is known that exchange-correlation functionals using the generalized gradient approximation (e.g. BLYP), although superior to the local density approximation, cannot describe long-range electron correlations that are responsible for van der Waals forces. These dispersion interactions ($\sim -C_6/r^6$) play a major role in simple fluids (e.g. argon) where they are responsible for the existence of a liquid phase. For complex systems like silicate melts, they compete with electrostatic and exchange-repulsion interactions although they are intrinsically much weaker than these ones. However, since the dispersion forces are always attractive they enhance the cohesive energy of the system (liquid or solid), a densification effect which could remedy to some of the inaccuracies encountered here with the use of GGA BLYP. Several routes (empirical or semi empirical) are proposed in the literature to estimate the dispersion contribution to the interaction energy (Becke and Johnson, 2005, 2007; Grimme, 2006; Civalleri et al., 2008): they are currently under investigation.

Despite these shortcomings, the AIMD calculation is clearly superior to the EPMD one to account for the intensity of the IR absorption bands in silicate melts, where the electronic polarization of the oxygen atoms is found to be the key mechanism in enhancing the dipolar absorption over a large frequency range. The ionic diffusivities deduced from the AIMD simulation seem to be in better agreement with the literature data than those calculated with the empirical potential model which tends to overestimate the ionic mobility especially in melts with high silica contents. This conclusion still needs to be confirmed by more extensive ab initio calculations. From the structural viewpoint, the AIMD calculations support the results obtained by EPMD (and by other classical simulation studies of the literature) namely the presence of a significant population of [5]Al in liquid silicates, the higher the number of NBO the higher the abundance of [5]Al. However, the concentration in [5]Al as evaluated by simulation is much larger than the one observed in aluminosilicate glasses for instance, a finding which points out the role played by the liquid state. This point certainly deserves to be investigated further by NMR or Raman spectroscopy.

Finally, it is noteworthy that AIMD calculations allow a detailed analysis of the electronic redistribution around the oxygen atoms as function of the local cationic environment and melt composition (this analysis will be the purpose of a forthcoming article). In a recent ab initio study (Salanne et al., 2008) devoted to the determination of the oxide ion polarizability in silicate melts, it has been shown that the distribution of the isotropic part of the oxygen polarizability evolves markedly with melt composition. Thus, the most probable value evolves from ~1.5 $A^3$ in silicic melt to ~1.8-2.0 $A^3$ in basic and ultrabasic melt, the distribution becoming broader when the degree of depolymerization of the melt increases. These findings could be helpful in giving a guideline to improve the empirical potentials for silicates. For instance, it is in a related framework that the aspherical ion model (Aguado et al., 2003; Madden et al., 2006; Jahn and Madden, 2007) has been developed to account for the polarizability and compressibility of the oxygen atom in oxide compounds. However, this model is still expensive in computer time (one or two order of magnitude greater than in classical simulations using empirical pair potentials) and we believe that, in taking into account the electronic redistribution around atoms by the way of ionic polarizabilities judiciously parametrized, a simple and accurate semi empirical potential model for silicate melts is close at hand.

**Acknowledgements**

We are grateful to C.J. Benmore and M.C. Wilding for their X-ray data on liquid silica and on magnesium silicate liquids. The anonymous reviewers are acknowledged for their helpful comments.

**BIBILIOGRAPHY**

**Table 1**

Chemical composition (in mol%) of the silicate melts simulated in this study. In parenthesis are the numbers of cations of each species used in the simulations (AIMD and EPMD) of the corresponding melt.

|  | $SiO_2$(mol%) | $Al_2O_3$(mol%) | MgO(mol%) | CaO(mol%) | $Na_2O$(mol%) | $K_2O$(wt%) |
|---|---|---|---|---|---|---|
| Silica | 100.0 (33) | 0.0 (0) | 0.0 (0) | 0.0 (0) | 0.0 (0) | 0.0 (0) |
| Rhyolite | 83.9 (26) | 8.1 (5) | 0.0 (0) | 0.0 (0) | 4.8 (3) | 3.2 (2) |
| $An_{36}Di_{64}$ | 50.0 (18) | 8.3 (6) | 16.7 (6) | 25.0 (9) | 0.0 (0) | 0.0 (0) |
| Enstatite | 50.0 (20) | 0.0 (0) | 50.0 (20) | 0.0 (0) | 0.0 (0) | 0.0 (0) |

**Table 2**

Structure of the simulated liquid silicates. For each ion is indicated: (first row) the first neighbor ion-oxygen mean distance in A and (second row) the average number of oxygens around the ion evaluated by integrating the first peak of the corresponding pair distribution function (the cut off distance corresponds to the location of the first minimum in the PDF, see Figs 2 and 3). The results of the ab initio calculation (AIMD) are listed in the first column and those obtained with the empirical potential (EPMD) are listed in the second column. A detailed discussion about the comparison with experimental data is given in the text.

| | | Si-O | | Al-O | | Mg-O | | Ca-O | | Na-O | | K-O | | O-O | |
|---|---|---|---|---|---|---|---|---|---|---|---|---|---|---|---|
| | | AIMD | EPMD | AIMD | EPMD | AIMD | EPMD | AIMD | EPMD | AIMD | EPMD | AIMD | EPMD | AIMD | EPMD |
| Silica | $R_{XO}$ | 1.65 | 1.63 | | | | | | | | | | | 2.70 | 2.65 |
| | N | 4.0 | 4.0 | | | | | | | | | | | 8.0 | 7.1 |
| Rhyolite | $R_{XO}$ | 1.65 | 1.64 | 1.77 | 1.73 | | | | | 2.40 | 2.46 | 2.80 | 3.20 | 2.70 | 2.66 |
| | N | 4.0 | 4.0 | 4.1 | 4.1 | | | | | 7.4 | 8.0 | 8.5 | 8.9 | 7.8 | 7.9 |
| $An_{36}Di_{64}$ | $R_{XO}$ | 1.64 | 1.63 | 1.77 | 1.74 | 2.0 | 2.0 | 2.50 | 2.38 | | | | | 2.71 | 2.66 |
| | N | 4.0 | 4.0 | 4.3 | 4.2 | 5.3 | 5.2 | 8.6 | 7.9 | | | | | 9.8 | 9.8 |
| Enstatite | $R_{XO}$ | 1.65 | 1.63 | | | 1.98 | 1.98 | | | | | | | 2.72 | 2.65 |
| | N | 4.0 | 4.0 | | | 5.2 | 5.2 | | | | | | | 11.8 | 10.6 |

**Table 3**

Population (in %) of n-coordinated cations in the four silicate melts simulated by AIMD and EPMD.

| | | Silica | | Rhyolite | | $An_{36}Di_{64}$ | | Enstatite | |
|---|---|---|---|---|---|---|---|---|---|
| | | AIMD | EPMD | AIMD | EPMD | AIMD | EPMD | AIMD | EPMD |
| $^{[n]}Si$ | n=3 | 1.0 | 0.2 | 2.1 | 0.7 | 1.2 | 0.9 | 0.9 | 1.5 |
| | n=4 | 98.6 | 99.6 | 96.9 | 97.8 | 94.7 | 94.3 | 95.5 | 93.8 |
| | n=5 | 0.4 | 0.2 | 1.0 | 1.5 | 4.1 | 4.7 | 3.4 | 4.6 |
| $^{[n]}Al$ | n=3 | | | 5.0 | 15.5 | 2.3 | 5.9 | | |
| | n=4 | | | 73.1 | 74.2 | 59.9 | 69.5 | | |
| | n=5 | | | 20.0 | 9.9 | 35.8 | 22.8 | | |
| | n=6 | | | 1.9 | 0.4 | 2.0 | 1.8 | | |
| $^{[n]}Mg$ | n=3 | | | | | 0.5 | 3.0 | 1.3 | 3.1 |
| | n=4 | | | | | 14.7 | 20.9 | 19.6 | 20.6 |
| | n=5 | | | | | 46.2 | 40.5 | 42.8 | 39.6 |
| | n=6 | | | | | 31.2 | 27.9 | 29.4 | 28.0 |
| | n=7 | | | | | 7.1 | 7.0 | 6.3 | 7.6 |
| | n=8 | | | | | 0.3 | 0.7 | 0.5 | 0.9 |
| $^{[n]}Ca$ | n=4 | | | | | 0.0 | 0.3 | | |
| | n=5 | | | | | 0.5 | 2.8 | | |
| | n=6 | | | | | 5.1 | 12.4 | | |
| | n=7 | | | | | 20.1 | 26.7 | | |
| | n=8 | | | | | 34.9 | 30.6 | | |
| | n=9 | | | | | 27.6 | 19.1 | | |
| | n=10 | | | | | 9.3 | 6.5 | | |
| | n=11 | | | | | 2.1 | 1.3 | | |
| | n=12 | | | | | 0.3 | 0.2 | | |
| $^{[n]}Na$ | n=3 | | | 0.0 | 1.8 | | | | |
| | n=4 | | | 1.3 | 5.7 | | | | |
| | n=5 | | | 4.4 | 12.7 | | | | |
| | n=6 | | | 18.3 | 19.8 | | | | |
| | n=7 | | | 29.5 | 21.8 | | | | |
| | n=8 | | | 26.9 | 18.2 | | | | |
| | n=9 | | | 13.8 | 11.8 | | | | |
| | n=10 | | | 4.4 | 5.4 | | | | |
| | n=11 | | | 1.4 | 1.7 | | | | |
| | n=12 | | | 0.0 | 0.4 | | | | |

**Table 4**

Populations (in %) of bridging oxygens (BO) and non-bridging oxygens (NBO) in the simulated melts. For each composition the first row corresponds to the ab initio results (AIMD) and the second row to those obtained with the empirical potential (EPMD). Notice that O-Si$_3$, Si$_2$-O-Al, Al$_2$-O-Si and O-Al$_3$ are triclusters.

| | | BO | | | | | | | | NBO | | | |
|---|---|---|---|---|---|---|---|---|---|---|---|---|---|
| | | Si-O-Si | Si-O-Al | Al-O-Al | O-Si$_3$ | Si$_2$-O-Al | Al$_2$-O-Si | O-Al$_3$ | $\sum N_{BO}/O_{Tot}$ | Si-O | Al-O | free oxygens | $\sum N_{NBO}/O_{Tot}$ |
| Silica | (AIMD) | 99.5 | | | 0.5 | | | | 100. | | | | |
| | (EPMD) | 99.9 | | | 0.1 | | | | 100. | | | | |
| Rhyolite | (AIMD) | 67.1 | 27.8 | 1.5 | 0.3 | 0.4 | 0.2 | 0.1 | 97.4 | 2.5 | 0.0 | 0.0 | 2.5 |
| | (EPMD) | 66.8 | 26.0 | 1.1 | 0.1 | 0.8 | 0.7 | 0.1 | 95.6 | 4.1 | 0.2 | 0.0 | 4.3 |
| An$_{36}$Di$_{64}$ | (AIMD) | 24.0 | 33.0 | 2.5 | 0.0 | 0.9 | 0.3 | 0.0 | 60.7 | 36.1 | 3.0 | 0.0 | 39.1 |
| | (EPMD) | 29.5 | 27.8 | 3.4 | 0.0 | 0.4 | 0.4 | 0.0 | 61.5 | 32.0 | 5.4 | 1.1 | 38.5 |
| Enstatite | (AIMD) | 34.0 | | | 0.0 | | | | 34.0 | 65.5 | | 0.4 | 65.9 |
| | (EPMD) | 38.5 | | | 0.1 | | | | 38.6 | 56.2 | | 5.0 | 61.2 |

**Table 5**

Ionic diffusivities in the silicate melts simulated with the empirical potential (EPMD) for the small system size (N~100 atoms, see text). The experimental values reported here for comparison (second, third and fourth row) are deduced from the experimental literature after a high temperature extrapolation based upon the Arrhenius plots given in the original papers.

| | T(K) | O | Si | Al | Mg | Ca | Na | K |
|---|---|---|---|---|---|---|---|---|
| | | | | $D_i$ ($10^{-9}$ m$^2$/s) | | | | |
| Silica | 2232 | $13.\times10^{-4}$<br>[a]$1.0\times10^{-5}$<br>[c]$6.1\times10^{-6}$ | $7.1\times10^{-4}$<br>[b]$1.5\times10^{-6}$ | | | | | |
| Rhyolite | 2243 | $7.1\times10^{-2}$<br>[d]$7.7\times10^{-3}$<br>[h]$\sim10^{-2}$ | $4.5\times10^{-2}$<br>[d]$7.1\times10^{-3}$<br>[i]$1.9\times10^{-3}$ | $0.13$<br>[e]$0.12$<br>[i]$4.6\times10^{-2}$ | | | $10.5$<br>[f]$25.8$<br>[g]$13.2$<br>[j]$11.8$ | $2.6$<br>[g]$1.2$ |
| An$_{36}$Di$_{64}$ | 2273 | $1.3$<br>[k]$0.4$<br>[l]$0.2$<br>[n]$0.3$ | $0.9$<br>[l]$0.2$<br>[n]$0.2$<br>[c]$0.45$ | $1.9$<br>[c]$0.9$ | $5.6$<br>[m]$2.1$ | $4.6$<br>[m]$1.7$<br>[o]$3.1$<br>[p]$0.6$ | | |
| Enstatite | 2573 | $2.6$<br>[q]$5.2$<br>[c]$2.7$ | $1.8$<br>[q]$5.2$<br>[r]$1.7$ | | $8.7$<br>[r]$5.5$ | | | |

[a] Mikkelsen (1984)

[b] Brebec et al. (1980)

[c] from the model of Mungall (2002) in using viscosity data by Urbain et al. (1982)

[d] O and Si in dacite by Tinker and Lesher (2001)

[e] Ga~Al in rhyolite by Baker (1992)

[f] Na in obsidian by Magaritz and Hofmann (1978)

[g] Na and K in obsidian by Jambon (1982)

[h] O in NaAlSi$_3$O$_8$ at 2100K by Poe et al. (1997)

[i] Si and Ga~Al in NaAlSi$_3$O$_8$ by Baker (1995)

[j] Na in pantellerite by Henderson et al. (1985)

[k] O in $An_{42}Di_{58}$ by Dunn (1982)

[l] O and Si in $An_{42}Di_{58}$ by Tinker et al. (2003)

[m] Mg and Ca in haplobasalt by La Tourette et al. (1996)

[n] O and Si in basalt by Lesher et al. (1996)

[o] Ca in basalt by Hofmann and Magaritz (1977)

[p] Ca in basalt by Jambon and Carron (1978)

[q] O and Si in diopside at 3GPa by Reid et al. (2001)

[r] Si and Mg in diopside at 1GPa by Shimizu and Kushiro (1991)

**Figures**

**Fig.1** Running average of the potential energy ($\Delta E_{pot} = <E_{pot} - \overline{E}_{pot}>$ where $\overline{E}_{pot}$ is the mean potential energy) and temperature fluctuations of the basic melt ($An_{36}Di_{64}$) simulated by AIMD as function of running time. The starting configuration of the AIMD calculation was provided by a well equilibrated EPMD simulation run. Notice that the equilibrium is reached after ~1.5 ps when $\Delta E_{pot}$ reaches a plateau value ($\Delta E_{pot} = 0$). Furthermore, the drift of the potential energy is very small on this timescale, a few kJ/mol, i.e. of the order of magnitude of the thermal fluctuations shown in the lower panel. For comparison, the corresponding quantities ($\Delta E_{pot}$ and T) are shown for the EPMD run starting from same initial configuration.

**Fig.2** Si-O, O-O and Si-Si pair distribution functions in the four simulated melts: AIMD simulations (full lines), EPMD simulations (dotted lines). For clarity, the curves are shifted vertically.

**Fig.3** As in Fig.2 but for other cation-oxygen and cation-cation pair distribution functions.

**Fig.4** X-ray radial distribution function for liquid silica at 2373K: AIMD simulation (full curve), EPMD simulation (dotted curves). The decomposition of the RDF in Si-O, Si-Si and O-O components is also shown. The experimental data (dots) are those obtained by Mei et al. (2007) at 2373K in using aerodynamic levitation and laser heating.

**Fig.5** As in Fig.4 but for liquid rhyolite. For clarity the atom-atom partial contributions associated with Na and K are not shown because they will be barely visible on the scale of the figure. In absence of experimental data for rhyolite, the data shown (dots) are those of Sugiyama et al. (1996) for liquid albite at 1460K (see text).

**Fig.6** As in Fig.4 but for the anorthite-diopside liquid mixture ($An_{36}Di_{64}$). For clarity, only the main ion-ion contributions are presented in the decomposition of the radial distribution function.

**Fig.7** As in Fig.4 but for liquid enstatite. The experimental data (dots) are those obtained by Wilding et al. (2008) in using aerodynamic levitation and laser heating.

**Fig.8** Population analysis of [n]-coordinated Si and Al as function of the cation-oxygen mean distance in rhyolite (dotted sticks), in $An_{36}Di_{64}$ (white sticks) and in enstatite (black sticks). The left figures show the EPMD results and the right figures those obtained by AIMD. The arrows indicate the [4]Si-O distance in simulated liquid silica.

**Fig.9** Population analysis of [n]-coordinated Mg and Ca as function of the cation-oxygen mean distance in $An_{36}Di_{64}$ (white sticks) and in enstatite (black sticks). The left figures show the EPMD results and the right figures those obtained by AIMD.

**Fig.10** Bond angle distributions in silica, rhyolite, $An_{36}Di_{64}$ and enstatite as obtained by AIMD (full curves) and EPMD (dotted curves).

**Fig.11** Log-log plot of the mean square displacement of atoms in the four simulated melts: AIMD simulations (full lines), EPMD simulations (dotted lines). For clarity, the behavior of the MSD calculated by EPMD are not shown below 1ps (10 ps for silica).

**Fig.12** Infrared absorption coefficient (in $cm^{-1}$) of liquid silica calculated by AIMD (full curve, with electronic polarization; dashed curve, without electronic polarization) and by EPMD (dotted curve, liquid; dashed-dotted curve, glass at 300K). For comparison is shown the absorption spectrum of vitreous silica (dots) by Velde and Couty (1987). For convenience, this experimental spectrum (given originally in arbitrary units) is normalized such as the maximum of the high frequency band at ~1100 $cm^{-1}$ equals 40,000 $cm^{-1}$ in the glassy state (as measured by Gervais et al., 1987) and 20,000 $cm^{-1}$ in liquid silica at 2000°C knowing that Markin and Sobolev (1960) observed a decrease of the intensity maximum of this band by a factor of two from the glass to the liquid.

**Fig.13** Infrared absorption coefficient (in $cm^{-1}$) of liquid rhyolite calculated by AIMD (full curve, with electronic polarization; dashed curve, without electronic polarization) and by EPMD (dotted curve). For comparison is shown the absorption spectrum of fused obsidian given in arbitrary units by Shimoda et al. (2004). For convenience, this experimental spectrum is normalized such as the maximum of the high frequency band is equal to the corresponding value reached by the AIMD calculation (i.e. ~16,500 $cm^{-1}$). Notice that in a sodium silica glass Gervais et al. (1987) evaluated the intensity maximum of the high frequency band to about 22,500 $cm^{-1}$.

**Fig.14** Partial absorption spectra associated with each ionic species present in the melt: silica (full curves), rhyolite (dotted curves), $An_{36}Di_{64}$ (dashed curves), and enstatite (dashed-dotted curves). The left panel shows the EPMD results and the right panel those obtained by AIMD (the electronic polarization is not accounted for in this case, see text). For clarity, the IR intensities associated with Al, Mg, Ca, Na and K are multiplied by an arbitrary factor (in parenthesis).

**Fig.15** Infrared absorption coefficient (in cm$^{-1}$) of liquid An$_{36}$Di$_{64}$ calculated by AIMD (full curve, with electronic polarization; dashed curve, without electronic polarization) and by EPMD (dotted curve). For comparison is shown the absorption spectrum of diopside glass (white dots) given in cm$^{-1}$ by Koike et al. (2000) and the reflectance spectrum of glassy basalt (full dots) given in arbitrary units by Minitti et al. (2002). For convenience, the reflectance spectrum of glassy basalt is normalized such as the absorption maximum of the high frequency band coincides with that of diopside glass.

**Fig.16** Infrared absorption coefficient (in cm$^{-1}$) of liquid enstatite calculated by AIMD (full curve, with electronic polarization; dashed curve, without electronic polarization) and by EPMD (dotted curve). For comparison is shown the absorption spectrum of enstatite glass (full dots) given in cm$^{-1}$ by Koike et al. (2000).

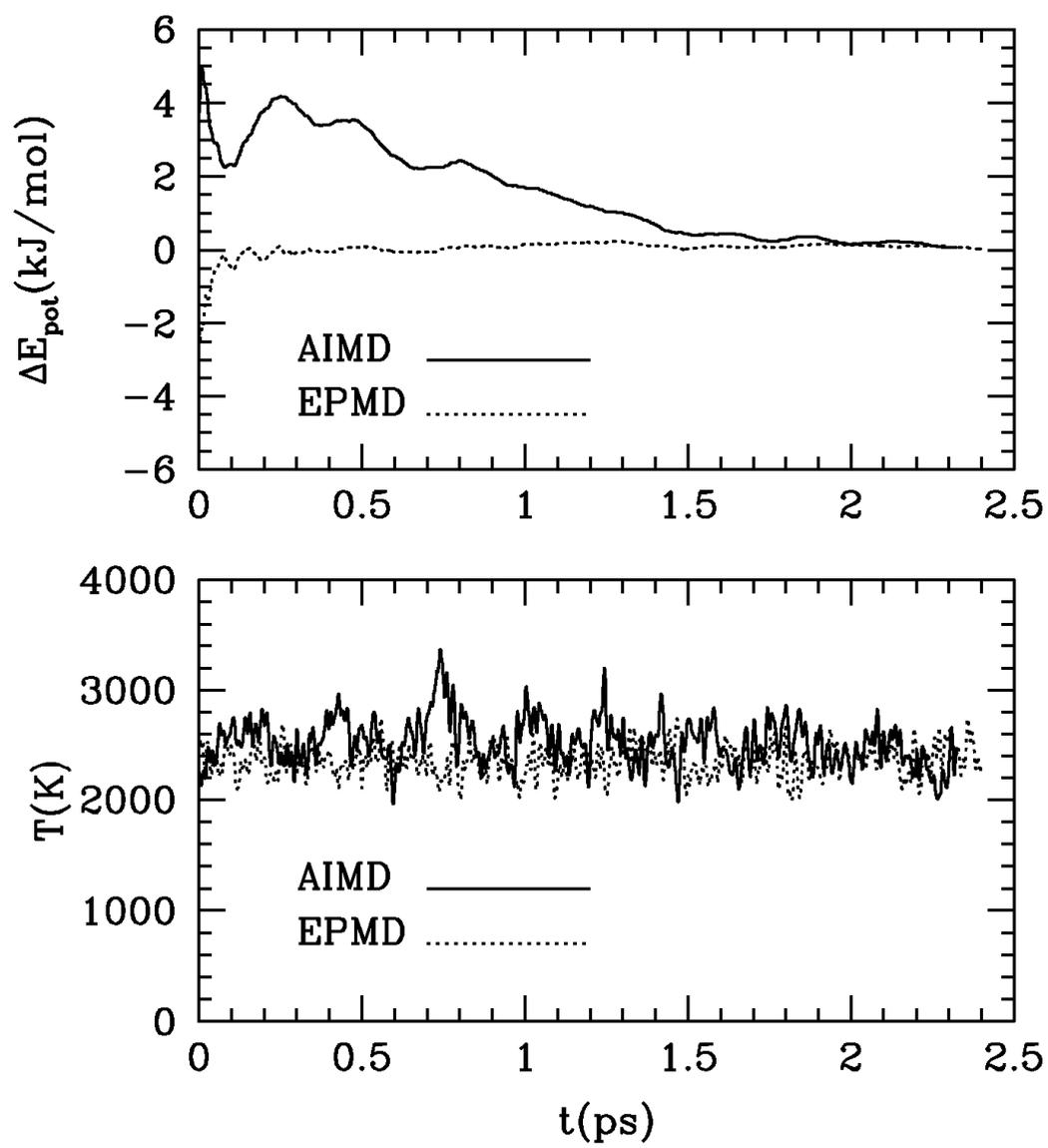

**Fig.1**

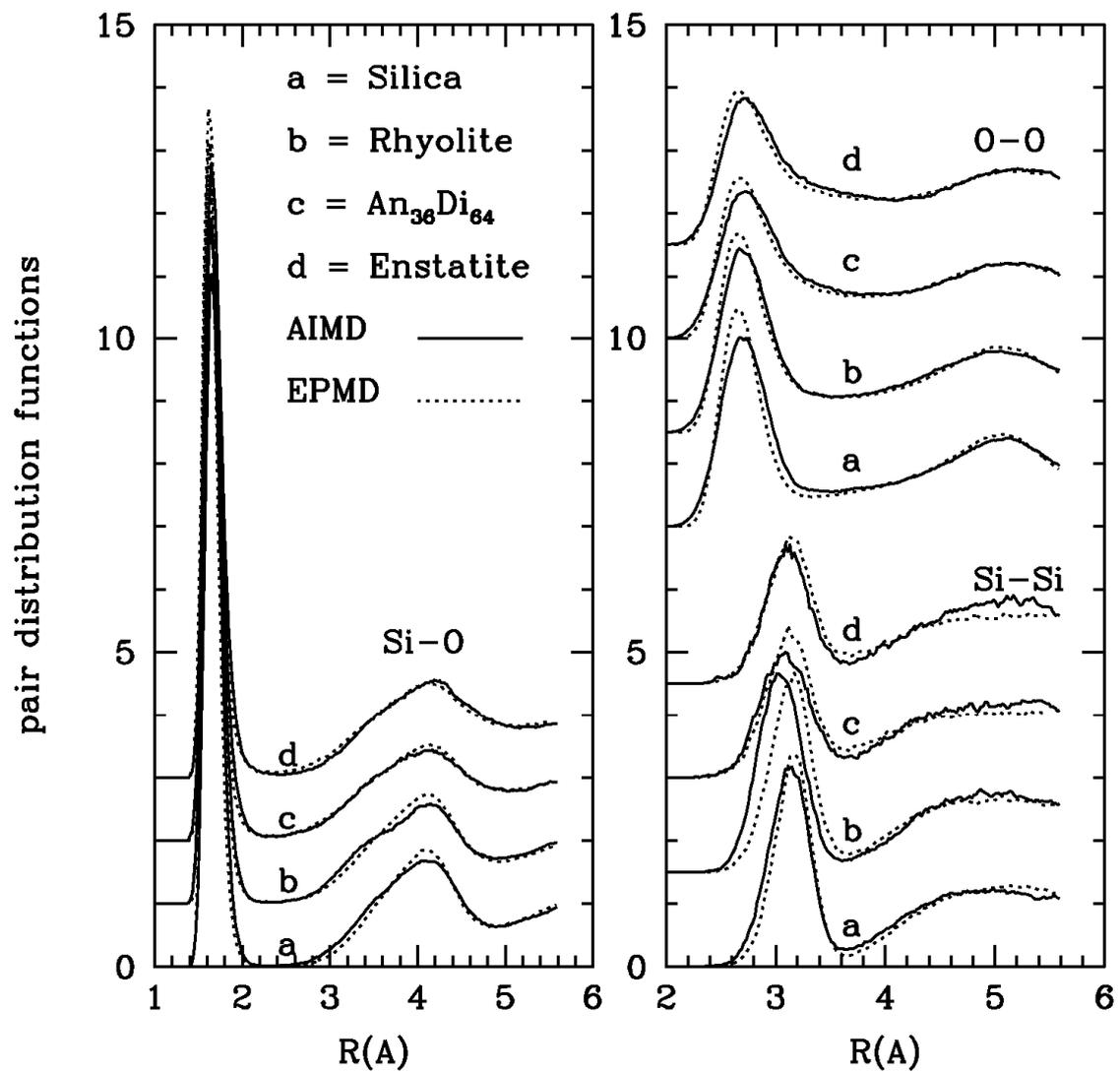

**Fig.2**

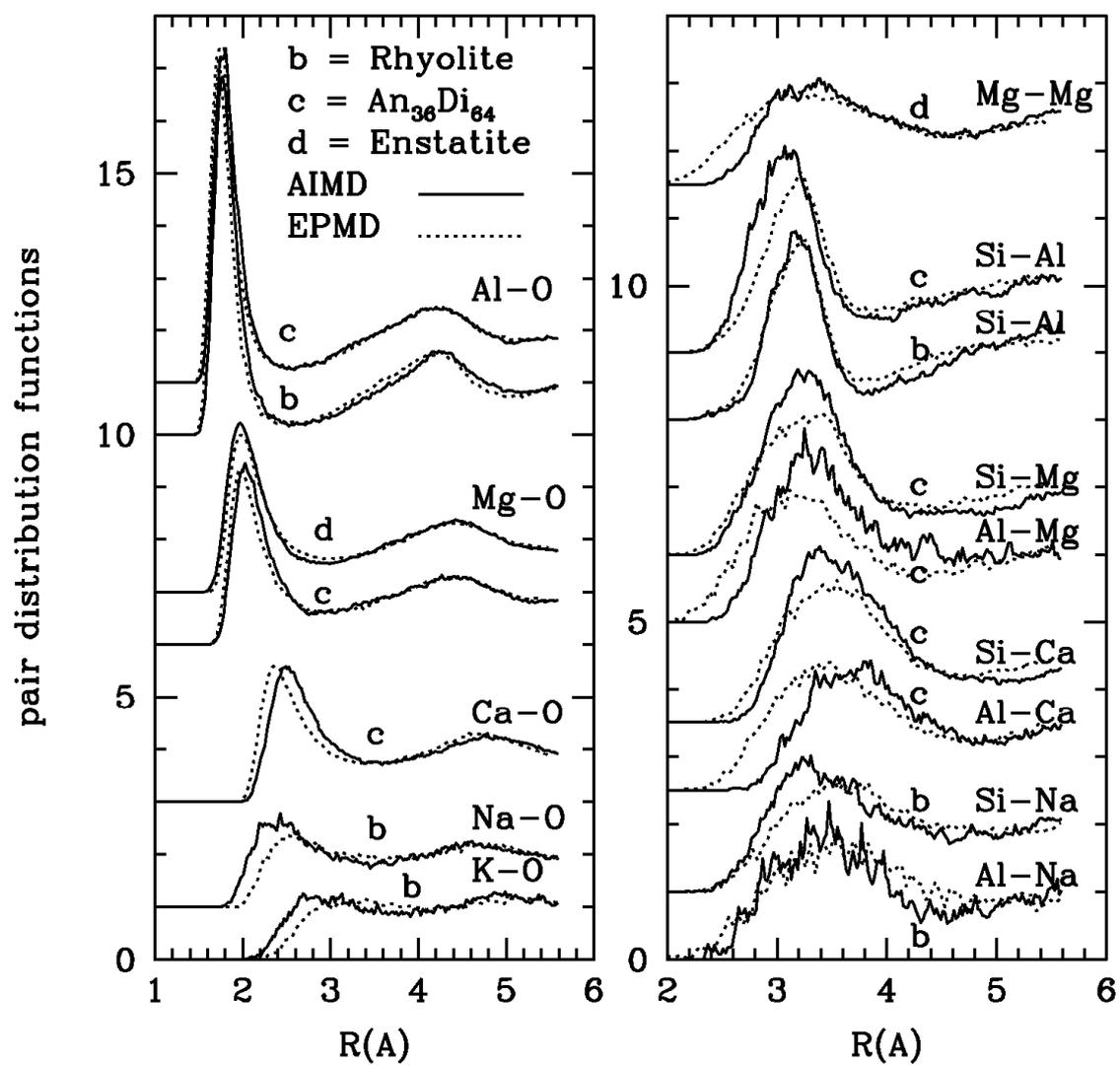

**Fig.3**

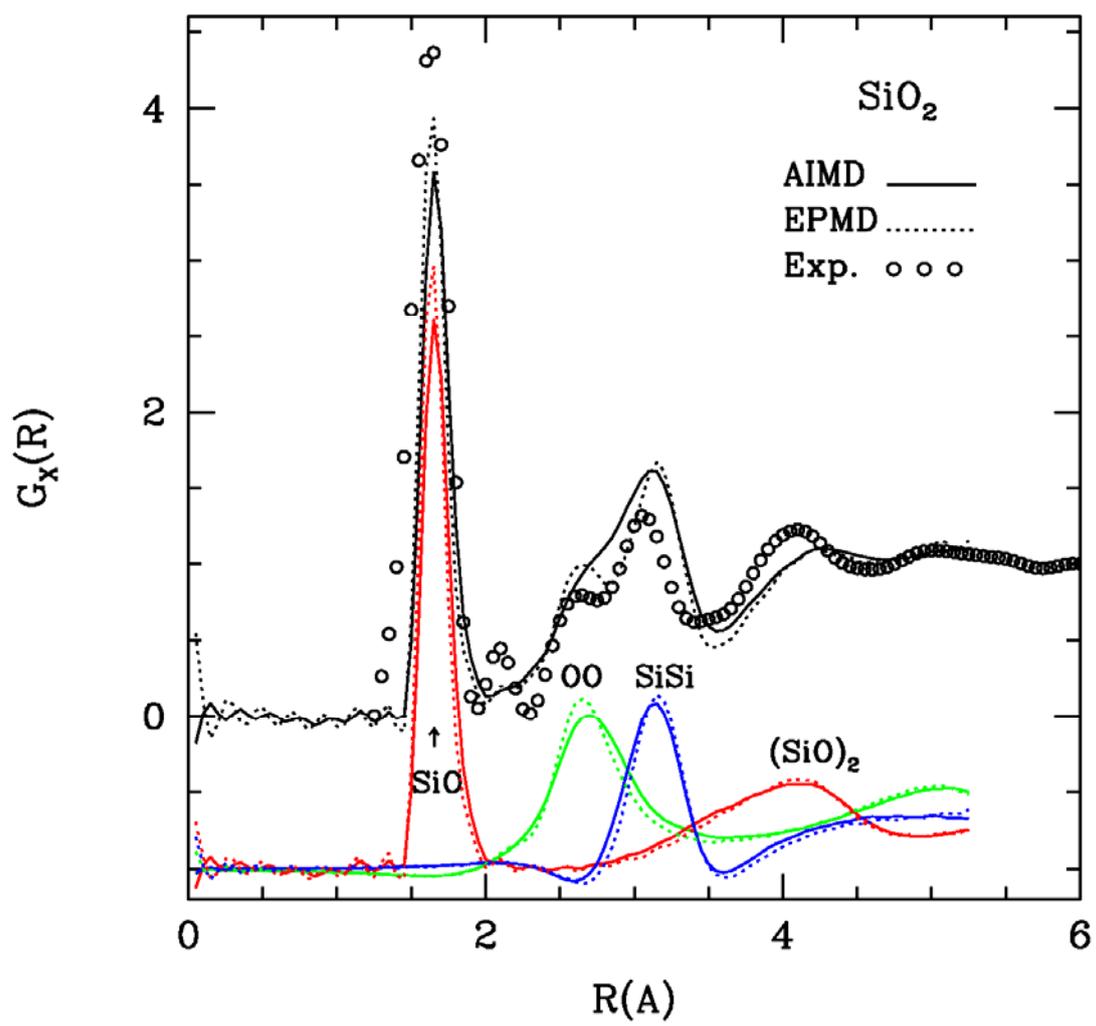

**Fig.4**

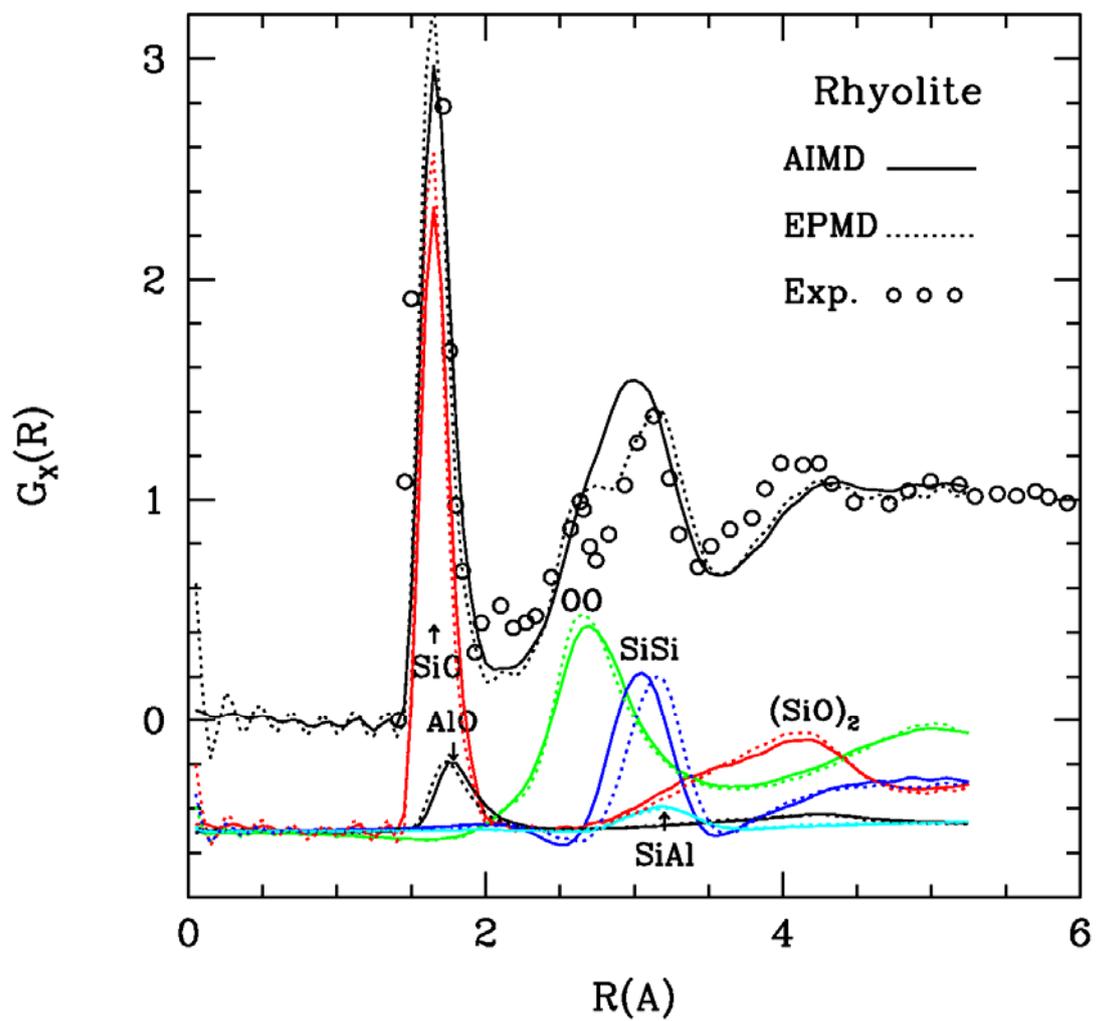

**Fig.5**

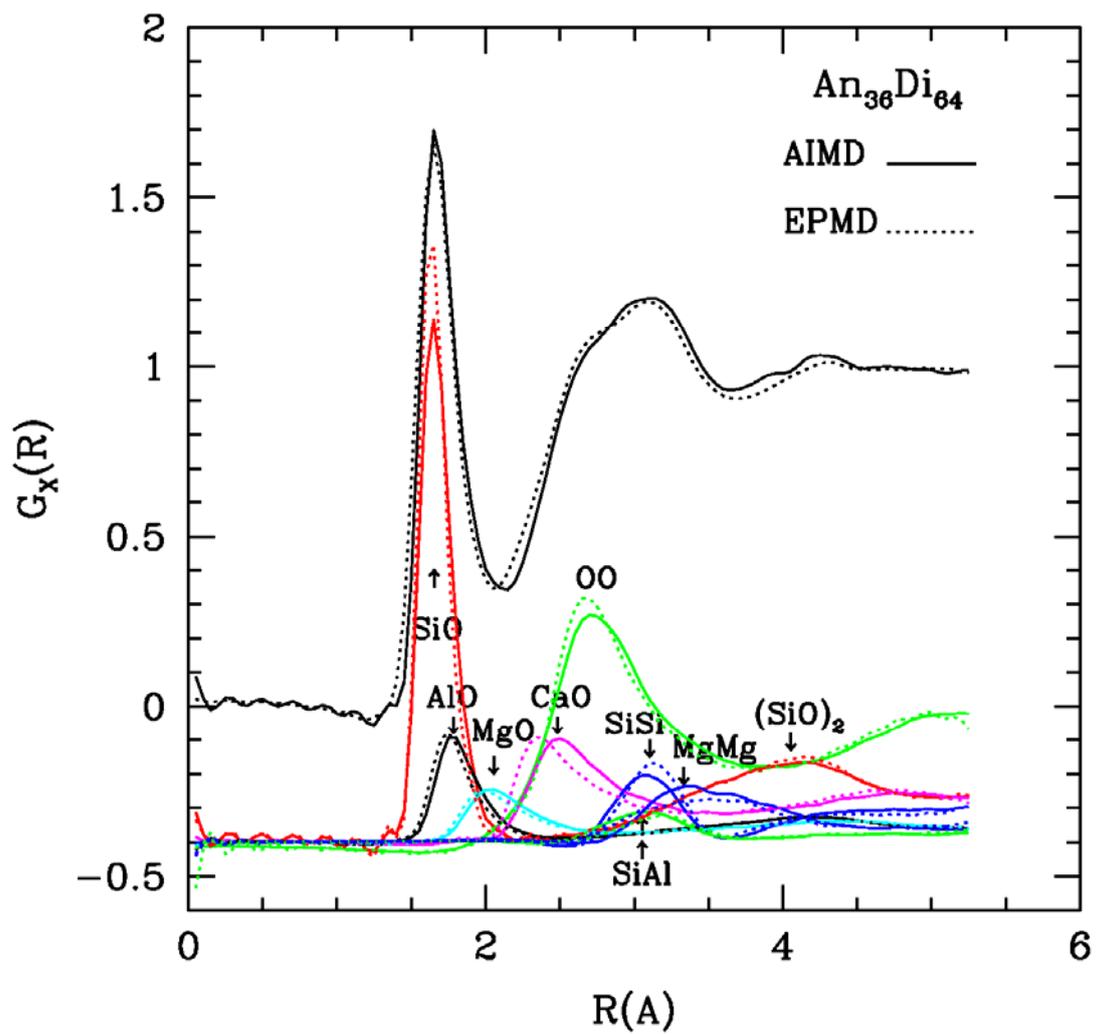

$An_{36}Di_{64}$

AIMD ———

EPMD ·········

OO

SiO

AlO

MgO

CaO

SiSi

MgMg

$(SiO)_2$

SiAl

**Fig.6**

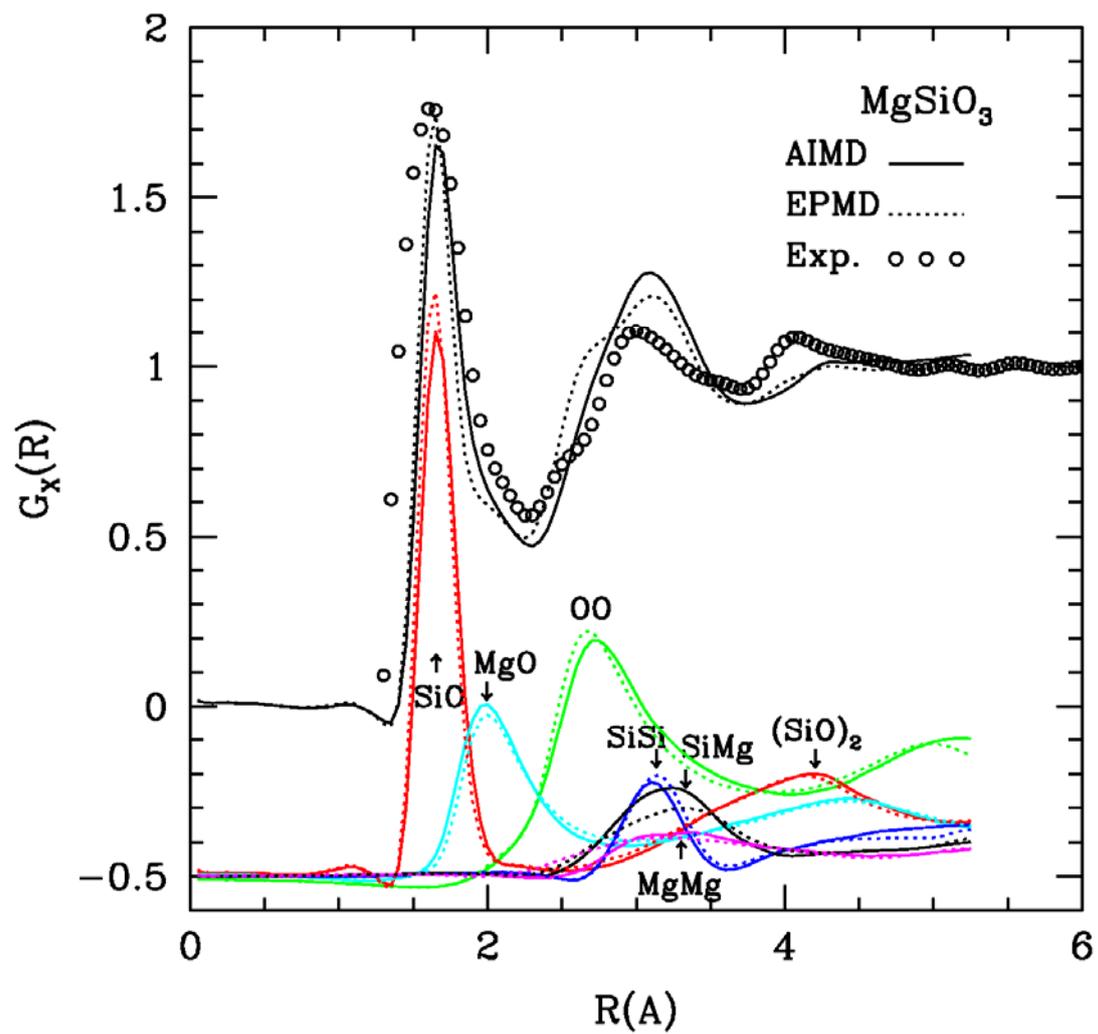

**Fig.7**

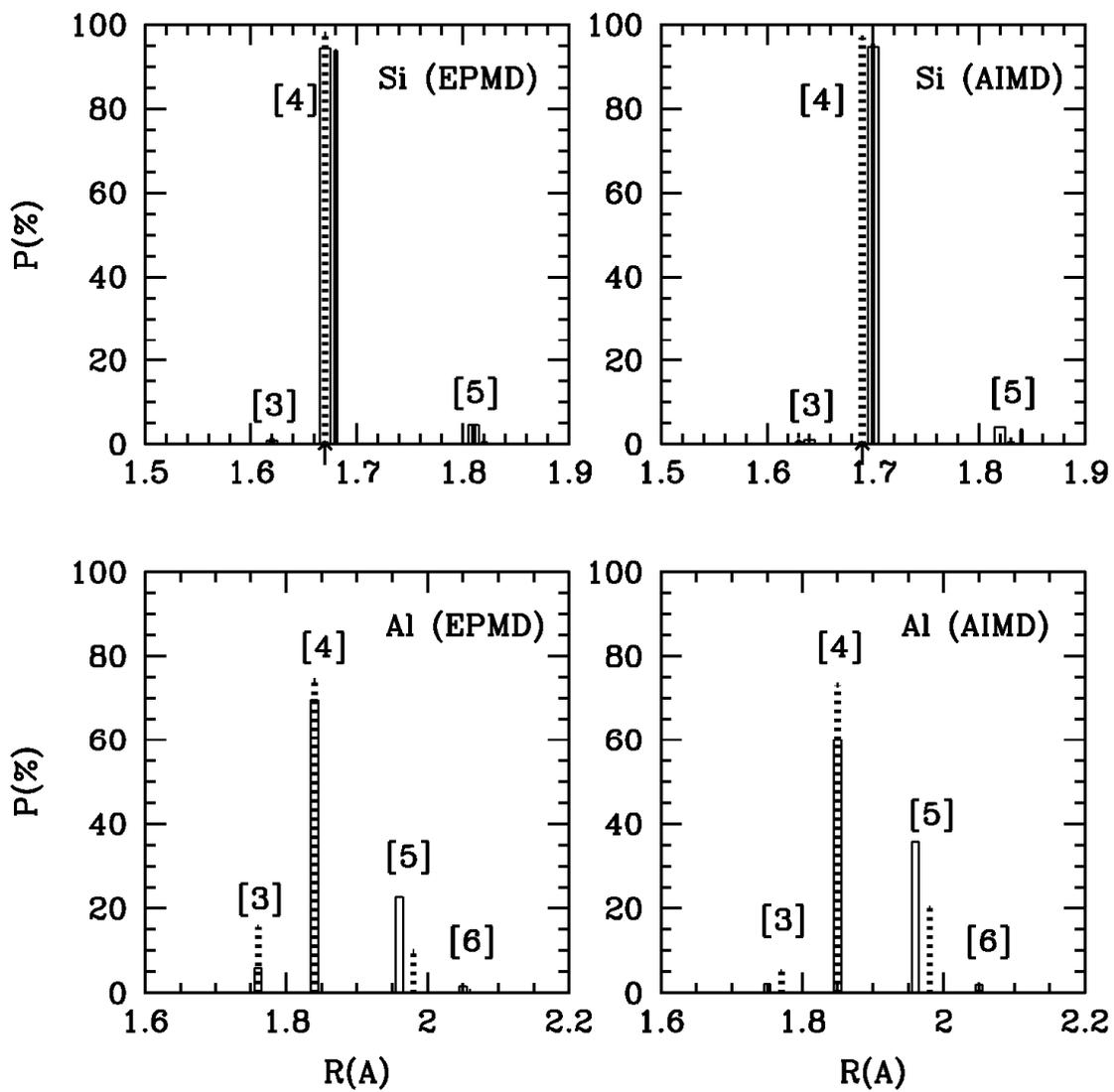

**Fig.8**

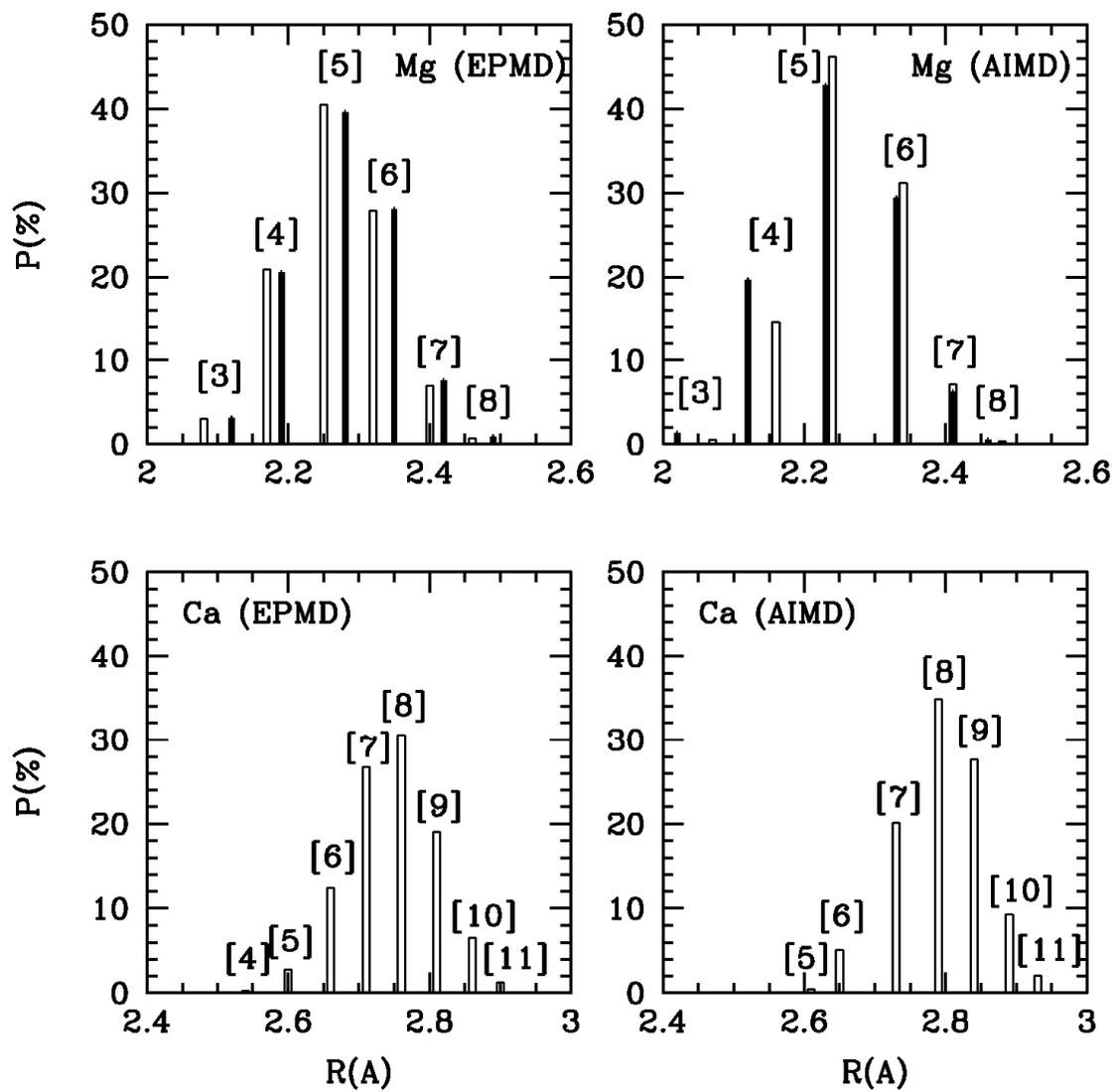

**Fig.9**

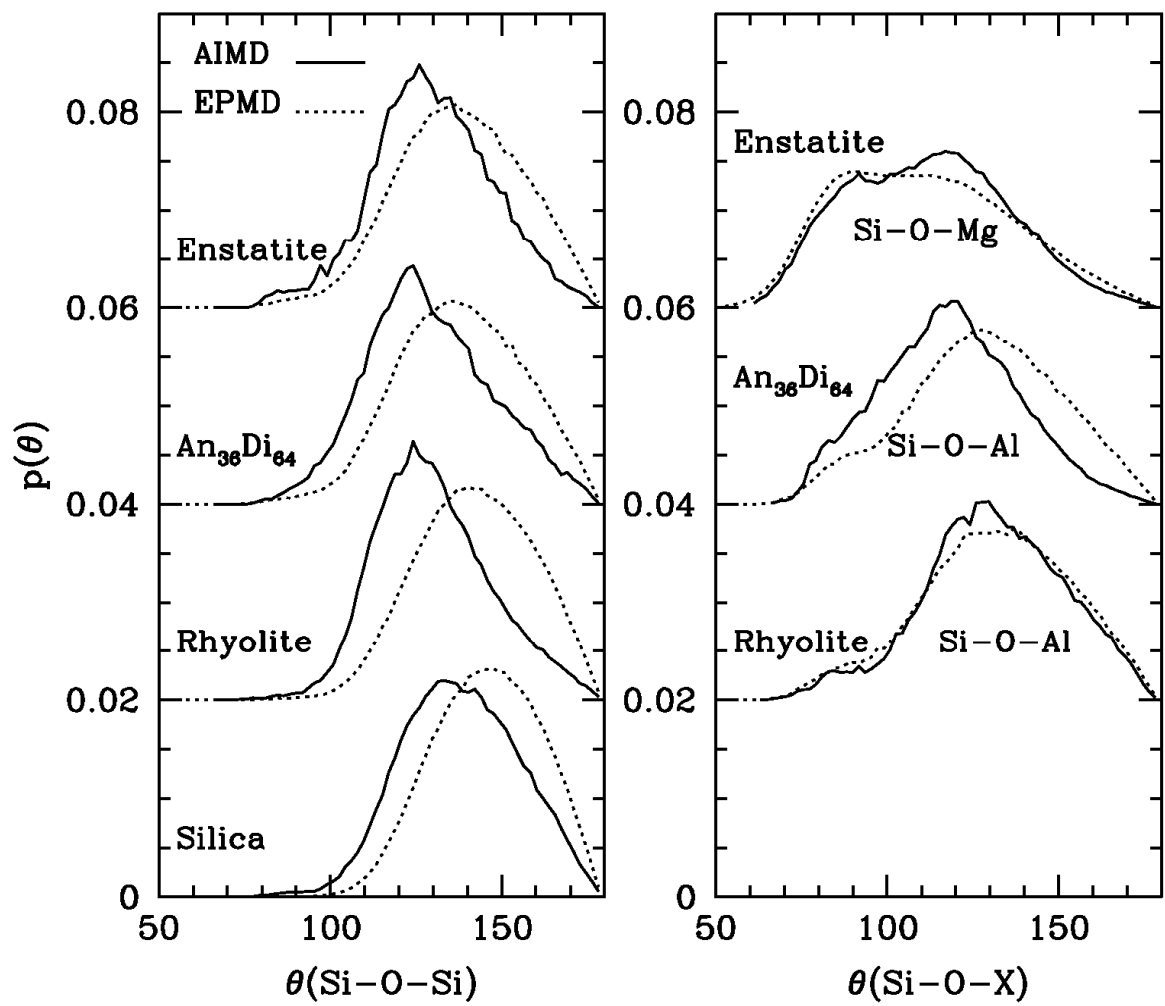

**Fig.10**

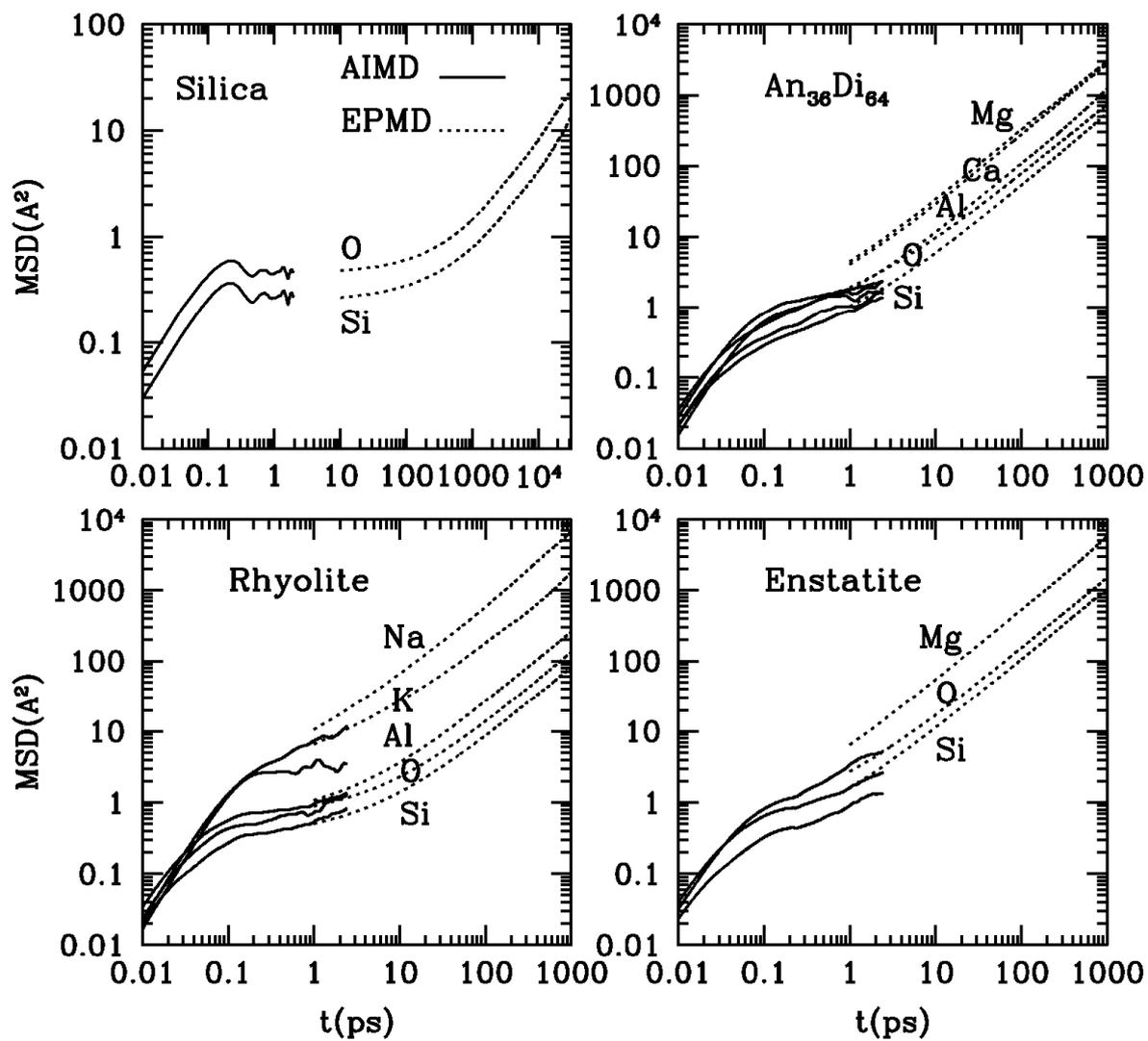

**Fig.11**

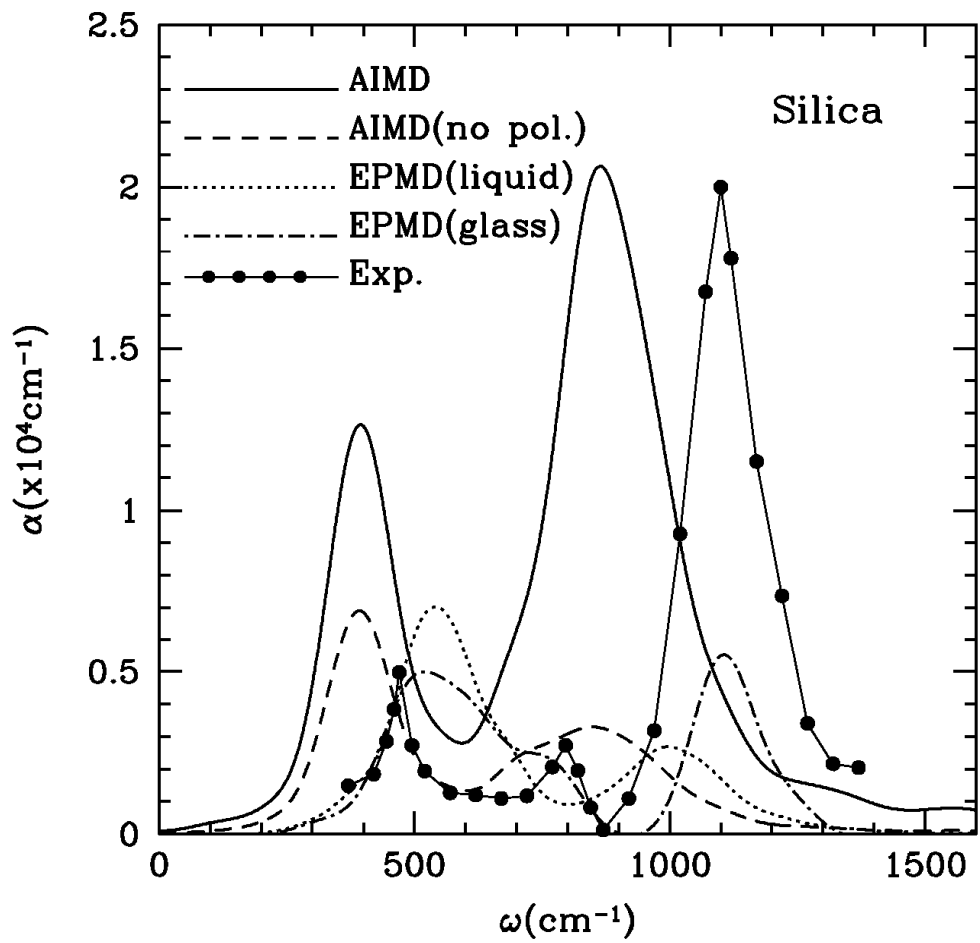

**Fig.12**

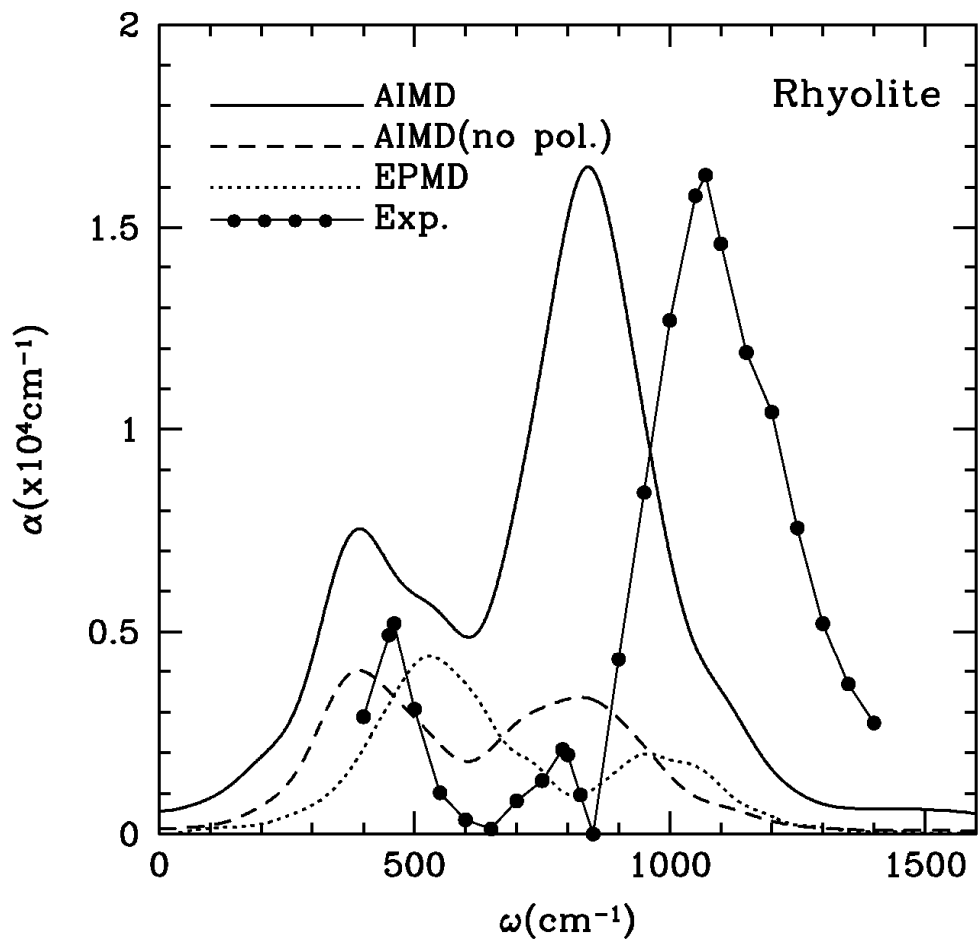

**Fig.13**

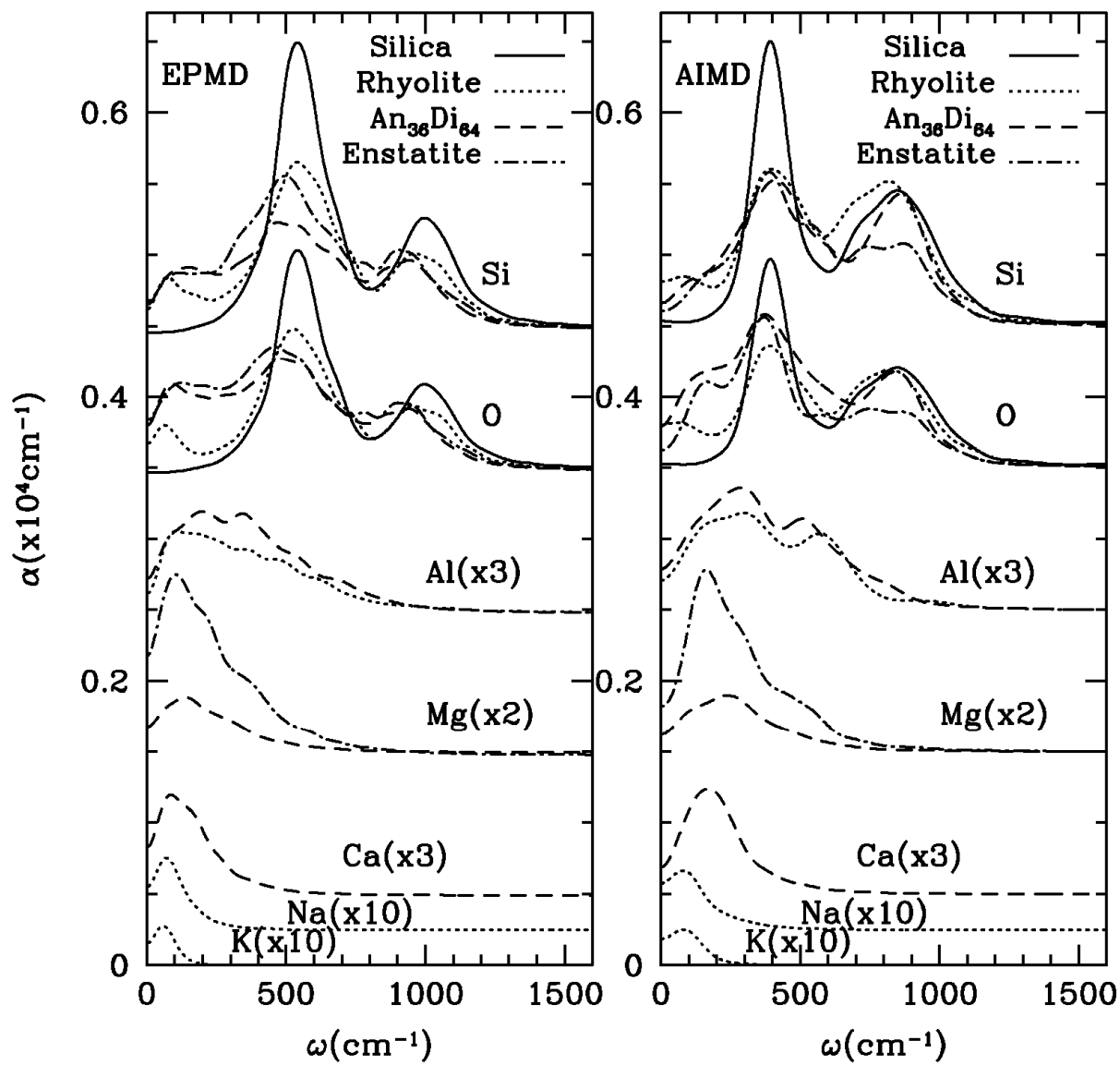

**Fig.14**

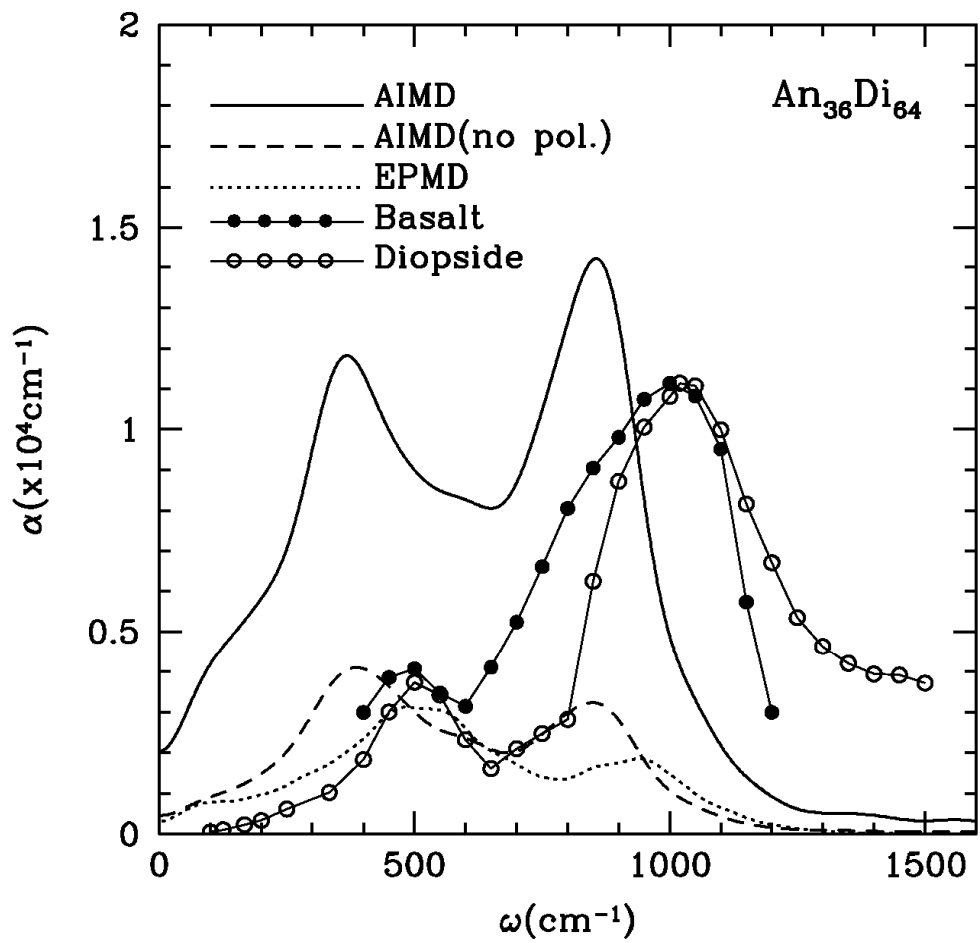

**Fig.15**

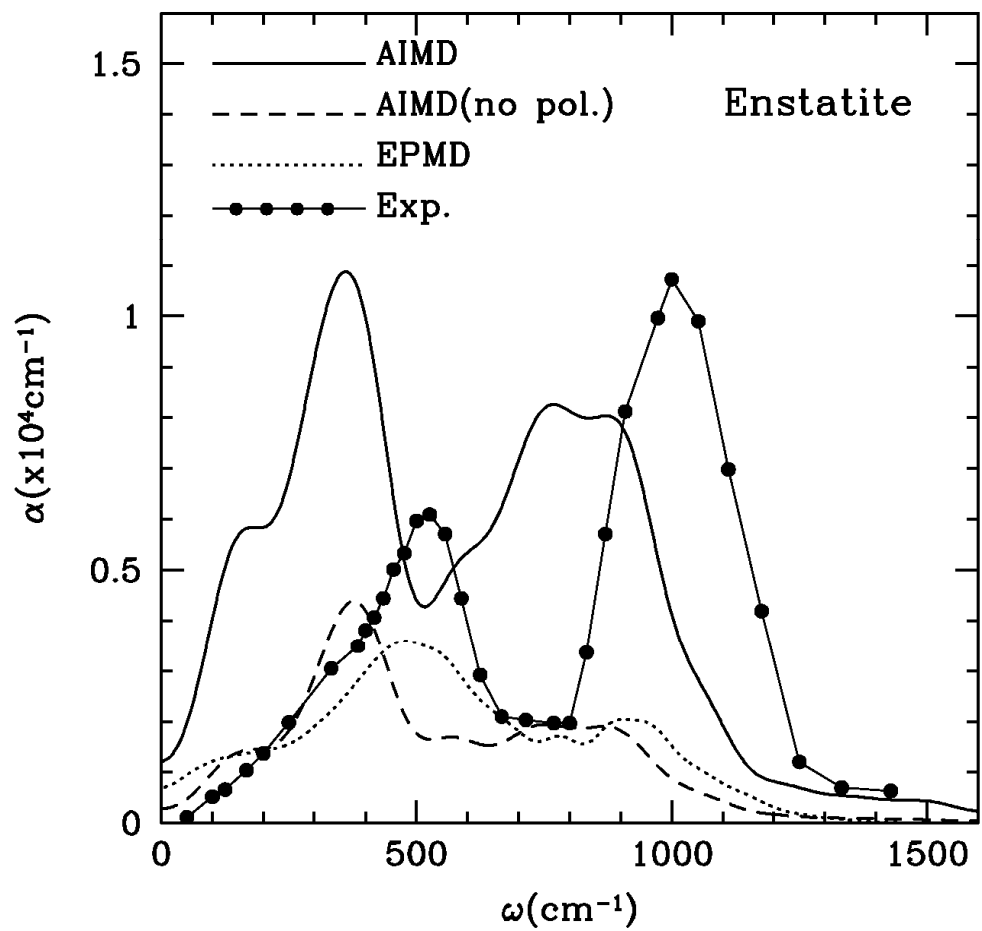

**Fig.16**